\documentclass[11pt]{article}
\usepackage{latexsym,amssymb,amsmath,times}
\usepackage{enumitem}
\usepackage{float}
\usepackage{hyperref}
\usepackage[square,sort&compress,comma,numbers]{natbib}

\usepackage{hyperref} 
\makeatletter
\@addtoreset{equation}{section} \makeatother

\input amssym.def
\input amssym.tex

\newcommand{\half}{{{\textstyle\frac{1}{2}}}}

\newcommand{\be}{\begin{equation}}
\newcommand{\ee}{\end{equation} }
\newcommand{\beqa}{\begin{eqnarray} }
\newcommand{\eeqa}{\end{eqnarray} }
\newcommand{\ba}{\begin{array}}
\newcommand{\ea}{\end{array}}

\newcommand{\Spin}{\mathbf{Spin}}
\newcommand{\GL}{\mathbf{GL}}

\newcommand{\rmd}{{\rm d}}

\newcommand{\ODD}{\mathbf{O}(D,D)}

\newcommand{\Ott}{\mathbf{O}(10,10)}

\newcommand{\Spint}{{\Spin(1,9)}}
\newcommand{\oSpint}{{{\Spin}(9,1)}}

\newcommand{\SLf}{\mathbf{SL}(5)}

\newcommand{\brM}{\bar{M}}

\newcommand{\DFT}{\rm{DFT}}

\newcommand\tr{{\rm tr\,}}

\newcommand\cA{{\cal A}}

\newcommand\cC{{\cal C}}

\newcommand\cF{{\cal F}}

\newcommand\cJ{{\cal J}}

\newcommand\cL{{\cal L}}
\newcommand\cM{{\cal M}}
\newcommand\cN{{\cal N}}

\newcommand\cP{{\cal P}}

\newcommand\cV{{\cal V}}

\newcommand\bcP{{\bar{\cP}}}

\newcommand\hcL{{\hat{\cal L}}}

\newcommand\dis{\displaystyle}

\def\tf{\tilde{f}}

\def\ty{\tilde{y}}

\def\bre{\bar{e}}

\def\breta{\bar{\eta}}
\def\bralpha{\bar{\alpha}}

\def\brp{{\bar{p}}}
\def\brq{{\bar{q}}}

\def\brF{\bar{F}}

\def\brK{\bar{K}}
\def\brL{\bar{L}}
\def\brR{\bar{R}}

\def\brV{{\bar{V}}}
\def\brW{{\bar{W}}}

\def\brP{{\bar{P}}}

\newcommand{\DO}{\mathbf{\nabla}}

\newcommand{\na}{{\nabla}}

\begin{document}
\begin{titlepage}
\title{\vskip -100pt
%\vskip 20pt
\vskip 2cm Comments on double field theory and diffeomorphisms \\}
\author{\sc{Jeong-Hyuck Park}}
\date{}
\maketitle \vspace{-1.0cm}
\begin{center}
Department of Applied Mathematics and Theoretical Physics, Cambridge, CB3 0WA, England\\
%~\\
{}\footnote{On sabbatical leave of absence for calendar year 2013.}Department of Physics, Sogang University, Mapo-gu,  Seoul 121-742, Korea\\
~\\
\texttt{park@sogang.ac.kr}
%~\\
~{}\\
%\texttt{imtak@,\quad kanghoon@sogang.ac.kr,\quad park@sogang.ac.kr,\quad yjsuh@sogang.ac.kr}
%\texttt{}\\
%{{\texttt{~{{{imtak@sogang.ac.kr\,,~kanghoon@sogang.ac.kr\,,~park@sogang.ac.kr}}}}}}
~~~\\~\\
\end{center}
\begin{abstract}
\vskip0.2cm
\noindent As the theory is   subject to a section condition,   coordinates in double field theory  do not   represent   physical points in an injective  manner. We argue that a  physical point  should      be rather    one-to-one   identified    with   a   `gauge orbit' in the coordinate space.   The diffeomorphism symmetry then  implies     an invariance under arbitrary reparametrizations of the gauge orbits. Within   this  generalized sense of  diffeomorphism,  we  show  that  a recently proposed  tensorial  transformation rule for finite coordinate transformations is actually    \textit{(i)}  consistent with the standard exponential map, and  further \textit{(ii)}  compatible with the full covariance of the `semi-covariant'  derivatives  and curvatures after projectors are properly imposed.      
\end{abstract}

%%%
{\small
\begin{flushleft}
~~%~\\
~~~~~~~~\textit{Keywords}:   Diffeomorphism, Double Field Theory, T-duality.
\end{flushleft}}
\thispagestyle{empty}
%%%%
%%%
%%02.40.-k 	Geometry, differential geometry, and topology
%%11.25.-w 	Strings and branes
%%04.60.Cf 	Gravitational aspects of string theory
%%04.65.+e 	Supergravity 
%%%
\end{titlepage}
\newpage
\tableofcontents %%
%%\begin{document} --> JHEP
%\twocolumn
%%%
%%\textbf{Introduction}. 
%%%
\section{Introduction}
\textbf{{Double Field Theory.}}    Ever since the inception of seminal ideas~\cite{Duff:1989tf,Tseytlin:1990nb,Tseytlin:1990va,Siegel:1993xq,Siegel:1993th}, through recent   renaissance revival~\cite{Hull:2009mi,Hull:2009zb,Hohm:2010jy,Hohm:2010pp},    there have been much   progress  in developing    T-duality manifest formalism  for  string theory effective actions,  especially    under the name, \textit{double field theory} (DFT)~\cite{Hull:2009mi,Hull:2009zb,Hohm:2010jy,Hohm:2010pp,Jeon:2010rw,Jeon:2011cn,Jeon:2011vx,Jeon:2012kd,Jeon:2011kp,Jeon:2011sq,Jeon:2012hp,Park:2012xn,Hohm:2010xe,Hohm:2011ex,Thompson:2011uw,Copland:2011yh,Hohm:2011dv,Albertsson:2011ux,Hohm:2011cp,Kan:2011vg,Geissbuhler:2011mx,Aldazabal:2011nj,Copland:2011wx,Hohm:2011nu,Copland:2012zz,Kan:2012nf,Grana:2012rr,Dibitetto:2012rk,Andriot:2012an,Copland:2012ra,Hohm:2011zr,Hohm:2011si,Hohm:2012gk,Dibitetto:2012xd,Hohm:2012mf,Aldazabal:2013mya,Geissbuhler:2013uka,Berman:2013uda}.  For analogous   parallel developments   on  U-duality in $\cM$-theory, we    refer to~\cite{Berman:2010is,Berman:2011pe,Berman:2011kg,Berman:2011cg,Berman:2011jh,Malek:2012pw,Berman:2012uy,Berman:2012vc,Musaev:2013rq,Park:2013gaj,Cederwall:2013naa,Cederwall:2013oaa,Godazgar:2013rja}.  Closely related  works based on   \textit{generalized geometry}~\cite{Hitchin:2004ut,Hitchin:2010qz,Gualtieri:2003dx}   are  notably \cite{Pacheco:2008ps,Grana:2008yw,Coimbra:2011nw,Coimbra:2011ky,Coimbra:2012yy,Coimbra:2012af,Koerber:2010bx}. \\

\noindent {DFT}   doubles    the spacetime dimension,  from $D$ to ${D+D}$, and     manifests   the $\ODD$  T-duality group structure.  The coordinates of the doubled spacetime, or \textit{DFT-coordinates},  $x^{A}=(\ty_{\mu},y^{\nu})$, may   decompose into  the ordinary coordinates, $y^{\mu}$,  and the dual `winding' coordinates, $\ty_{\nu}$. The doubling   also reflects the existence of  the left and right modes in closed strings.  Yet,   DFT  ought to be  a $D$-dimensional theory  formulated in a doubled spacetime.    Being subject to    so called  the \textit{strong constraint} or \textit{section condition},  DFT may reduce to a usual    string theory effective action in $D$-dimension.  The section condition requires that DFT lives on a $D$-dimensional null hyperplane, such that  the $\ODD$ invariant d'Alembertian operator is  trivial,%\footnote{The $\ODD$ vector indices (capital roman, $A,B,C,\dots$) are raised or lowered by the $\ODD$ metric, $\cJ$ in (\ref{ODDmetric}).}
\be
\ba{ll}
\partial_{A}\partial^{A}=\cJ^{-1AB}\partial_{A}\partial_{B}\equiv 0\,,\quad&\quad\cJ_{AB}=\left(\ba{cc}0&1\\1&0\ea\right)\,,
\ea
\label{ODDmetric}
\ee
acting on arbitrary fields  as well as their products,
\be
\ba{ll}
\partial_{A}\partial^{A}\Phi(x)=0\,,\quad&\quad
\partial_{A}\Phi_{1}(x)\partial^{A}\Phi_{2}(x)=0\,.
\label{seccon}
\ea
\ee
Further, the \textit{DFT-diffeomorphism symmetry}   is  generated by   a  generalized Lie derivative,
\be
\hcL_{V}T_{A_{1}\cdots A_{n}}:=V^{B}\partial_{B}T_{A_{1}\cdots A_{n}}+\omega\partial_{B}V^{B}T_{A_{1}\cdots A_{n}}+\sum_{i=1}^{n}(\partial_{A_{i}}V_{B}-\partial_{B}V_{A_{i}})T_{A_{1}\cdots A_{i-1}}{}^{B}{}_{A_{i+1}\cdots  A_{n}}\,,
\label{tcL}
\ee
where   $\omega$ is the weight of the DFT-tensor,  $T_{A_{1}\cdots A_{n}}(x)$,   and  $V^{A}(x)$ corresponds to the \textit{infinitesimal DFT-diffeomorphism  parameter} which also obeys  the section condition,
\be
\ba{ll}
\partial_{A}\partial^{A}V^{B}(x)=0\,,\quad&\quad \partial_{A}V^{B}(x)\partial^{A}\Phi(x)=0\,.
\ea
\label{secconV}
\ee
~\\
\noindent The pioneering   works on  DFT~\cite{Hull:2009mi,Hull:2009zb,Hohm:2010jy,Hohm:2010pp} focused on   the  NS-NS sector. The action was written 
 in terms of `ordinary' derivatives  acting on a ``generalized  metric". While the $\ODD$ structure was manifest,  the diffeomorphism symmetry~(\ref{tcL}) was rather hidden.  Since then, to the best of our knowledge, there have been three kinds of proposals  for the underlying differential geometry of DFT,   apart from the (undoubled) generalized geometry approach~\cite{Coimbra:2011nw,Coimbra:2011ky,Coimbra:2012yy,Coimbra:2012af}.  \\
\indent Firstly,   involving the present author,  a \textit{semi-covariant} derivative was introduced,   to start with     for the NS-NS sector~\cite{Jeon:2010rw,Jeon:2011cn},  and     generalized   further to   fermions~\cite{Jeon:2011vx},    to the   {R-R} sector~\cite{Jeon:2012kd} as well as   to    Yang-Mills~\cite{Jeon:2011kp}.    The crucial  feature of the semi-covariant derivatives   is that,  combined with   `projection' operators,   they can be made   fully covariant.   In this approach which we henceforth refer to  as \textit{T-geometry} \textit{c.f.~}\cite{Park:2013gaj},  the original ``generalized  metric"  was   traded by a pair of orthogonal and complete  projectors,  
\be
\ba{llll}
P_{A}{}^{B}P_{B}{}^{C}=P_{A}{}^{C}\,,~~~~&~~~~\brP_{A}{}^{B}\brP_{B}{}^{C}=\brP_{A}{}^{C}\,,\quad&\quad
P_{A}{}^{B}\brP_{B}{}^{C}=0\,,~~~~&~~~~P_{A}{}^{B}+\brP_{A}{}^{B}=\delta_{A}{}^{B}\,,
\ea
\label{projection}
\ee
such that   a conceptual   emphasis was  put   on the \textit{action  of   projection}  (into two opposite modules),  rather than  questioning  any geometric meaning of the generalized metric, \textit{e.g.~}``measuring   length".\footnote{In fact, in the full order supersymmetric extensions of ${D=10}$ DFT,  both   ${\cN=1}$~\cite{Jeon:2011sq} and    ${\cN=2}$~\cite{Jeon:2012hp}, the $1.5$ formalism which is familiar in supergravities    works  nicely with  the projectors  rather  than with  the generalized metric,  \textit{c.f.} footnote~\ref{foot1p5}.}  After all, it is   the  flat  $\ODD$ metric~(\ref{ODDmetric}) that  raises or lowers the positions of  the DFT  vector indices.\\
\indent Secondly,  based   on an earlier work~\cite{Siegel:1993xq},   more orthodox approach was pursed to postulate a  fully  covariant derivative~\cite{Hohm:2011si,Hohm:2012gk}. Yet, the required  connection with all the desired  properties  turned out to inevitably contain  non-physical parts  which cannot  be constructed from the NS-NS sector.   The undecidable  non-physical parts amount  to  the $(2,1)$  traceless  $\GL(D)$  Young tableau.  After `projecting' them  out, this    approach  was shown to  be consistent   with  the aforementioned  semi-covariant,  T-geometry~\cite{Hohm:2011si}.     An   index-free, basis-independent   formulation of  this  approach has been also  addressed   in \cite{Hohm:2012mf} along with      in-depth  discussion on    curvatures.   \\
\indent Thirdly (and very recently), another  interesting   approach was proposed in \cite{Berman:2013uda} where the  Weitzenb\"{o}ck connection was employed which, compared to the previous two approaches,  assumes a simple form. It leads to a fully covariant   derivative for the diffeomorphism symmetry, but breaks the  local Lorentz symmetry. Demanding the local Lorentz  symmetry  at the whole action level,  DFT can  be restored, at least for the NS-NS sector.   One novel feature in this approach is that ---quite opposite to  the T-geometry attitude---   the resulting action does not contain  the flat  $\ODD$ metric at all, as  the $\ODD$ vector indices are there lowered and  raised   by the generalized metric and its inverse. The $\ODD$ structure only arises when specifying the generalized metric. This approach  then  opens up a  possibility to  establish  a direct link between  DFT and   U-duality manifest $\cM$-theory effective actions~\cite{Thompson:2011uw,Berman:2011pe,Berman:2011kg,Berman:2011cg,Malek:2012pw,Berman:2012uy,Berman:2012vc,Musaev:2013rq,Park:2013gaj,Cederwall:2013naa,Cederwall:2013oaa,Godazgar:2013rja,Berman:2013uda} where there appears  only a  generalized metric  and no  extra flat metric.\\

\noindent While all the  three approaches  seem to complement to each other, it is also true    that  so far  the full order supersymmetric completions  of  DFT have been accomplished    within the semi-covariant T-geometry setup only,    as   ${\cN=1}$ ${D=10}$~\cite{Jeon:2011sq} and    ${\cN=2}$ ${D=10}$~\cite{Jeon:2012hp}.\footnote{\textit{c.f.~}\cite{Coimbra:2011nw,Coimbra:2011ky,Coimbra:2012yy,Coimbra:2012af,Hohm:2011nu} for   linear order  analysis with different details.}   The full order supersymmetric completion may be viewed as  a direct  proof of the \textit{existence} of the supersymmetric double field theory (SDFT).  Especially the $\cN=2$ ${D=10}$ SDFT  unifies type IIA and IIB supergravities in a  manifestly covariant manner  with respect to  all  the  bosonic symmetries listed in  Table~\ref{TABsymmetry}.  While the theory  is unique, the solutions turn out to be  twofold. Type  IIA and IIB supergravities are   identified there  as two different types of solutions rather than two different theories.\\
\begin{table}[H]
\begin{center}
\begin{itemize}
\item \textit{$\Ott$ T-duality~}  (hidden symmetry)
\item \textit{DFT-diffeomorphisms,  generated  by the  generalized Lie derivative}~(\ref{tcL})
\item \textit{A pair of local    Lorentz  symmetries,} ${\Spint\times\oSpint}$
\item ${\cN=2}$ \textit{ local supersymmetry  of 32 supercharges }
\end{itemize}
\caption{{Symmetries of $\cN=2$ $D=10$  SDFT.~}      
 The doubling of the Lorenz groups  reflects  the existence of the   left and  right  modes in closed strings. The $\Ott$ T-duality is \textit{a priori}     not   a Noether symmetry.  }
\label{TABsymmetry}
\end{center}
\end{table}
%%%%%%%%%%%%%%%%%%%%%%%%%%%%%%%%%%%%%%%%%%%%%%%%%%%%%%%%%%%%%%%%%%%%%%%%%%%%%%%%%%%%%%%%%%%%%%%%%%%%%%%%%%%%%%%%%%%%%%%%%%%%%%%%%%%%%%%%%%%%%%%%%%%%%%
\noindent\textbf{{Beyond  Supergravity.}}    The    aspects of DFT   depicted above mainly  concern   the theme of reformulating (or unifying) the  already known  supergravities in a duality manifest manner. Yet,  there are also evidences that DFT may well  extend   beyond the supergravity realm.   Applying the  Scherk-Schwarz reduction method to DFT,  it has been realized  that one can actually relax the section  condition~(\ref{seccon})~\cite{Geissbuhler:2011mx} ---and hence beyond supergravity--- and derive  all the known gauged supergravities in lower than ten dimensions~\cite{Geissbuhler:2011mx,Aldazabal:2011nj,Grana:2012rr,Dibitetto:2012rk,Dibitetto:2012xd,Aldazabal:2013mya,Geissbuhler:2013uka,Musaev:2013rq,Berman:2013uda}.\\
\indent Further reasons to believe that DFT is  not  a mere rewriting  of  supergravities are, in our view, threefold:   \textit{(i)} the very usage of $\ODD$-covariant genuine    DFT-field variables,  alternative to the Riemannian ones,  such as metric and  form-fields,   \textit{(ii)} the doubling of the local Lorenz symmetry,\footnote{This property is also shared  with  generalized geometry~\cite{,Coimbra:2011nw,Coimbra:2011ky,Coimbra:2012yy,Coimbra:2012af}.} ${\Spint\times\oSpint}$,  and  \textit{(iii)} the possibility of  combining  DFT-diffeomorphism   with $\ODD$ rotations. \\
\indent  Historically, the generalized metric was  spelled  at first   as a composite of the metric and the $B$-field~\cite{Duff:1989tf}. Yet, it has become  evident that it  is also  possible to \textit{define} the generalized metric  in a more abstract and covariant fashion,  \textit{i.e.~}simply as  a symmetric $\ODD$ element, since  the most general form of such a ${2D\times 2D}$ matrix can be parametrized  or solved  by a pair of ${D\times D}$ matrices,   one symmetric and the other  anti-symmetric.  Naturally these  can be identified as the  Riemannian metric and the $B$-field.   However, the parametrization is not unique: $\ODD$ rotations as well as various field redefinitions can  be freely applied.\footnote{The parametrization is only unique up to the $\ODD$ rotations and field redefinitions.} Only a concrete  choice of the section, \textit{e.g.~}$\frac{\partial~~}{\partial \tilde{y}_{\mu}}\equiv 0$, may pin down a specific   parametrization to be compatible with the  standard Riemannian geometry. However,  when  we consider  further reductions  from $D$ to lower dimensions, there is no longer a single  preferred parametrization. It may well be  better  to work with  the parametrization-independent and  $\ODD$ covariant genuine DFT-field variables without taking any parametrization, at least for the compactified sector if not all.\\  
\indent Especially  in the construction of the ${\cN=2}$ ${D=10}$ SDFT~\cite{Jeon:2012hp},  in order  to   manifest   all the   symmetric structures, and also  for the successful full order supersymmetric completion, it was necessary to postulate   the fundamental fields     to be precisely the following variables:   The DFT-dilaton,  $d$,  ~DFT-vielbeins, $V_{Ap}$, $\brV_{A\brp}$,  ~and the  R-R potential,  $\cC^{\alpha}{}_{\bralpha}$, ~plus fermions.  The DFT-vielbeins are \textit{defined} to  satisfy  the following four   algebraic relations~\cite{Jeon:2011cn,Jeon:2011vx},\footnote{In (\ref{defV}), ${\eta_{pq}=\mbox{diag}(-++\cdots+)}\,$ and $\,{\breta_{\brp\brq}=\mbox{diag}(+--\cdots-)}\,$ are  ten-dimensional  metrics for $\Spint$ and $\oSpint$  respectively~\cite{Jeon:2012kd,Jeon:2012hp}. }
\be
\ba{llll}
V_{Ap}V^{A}{}_{q}=\eta_{pq}\,,~~~&~~~
\brV_{A\brp}\brV^{A}{}_{\brq}=\breta_{\brp\brq}\,,~~~&~~~
V_{Ap}\brV^{A}{}_{\brq}=0\,,~~~&~~~ V_{Ap}V_{B}{}^{p}+\brV_{A\brp}\brV_{B}{}^{\brp}=\cJ_{AB}\,,
\ea
\label{defV}
\ee
such that they generate   the  pair of  projectors~(\ref{projection}) as  $P_{AB}=V_{A}{}^{p}V_{Bp}$, $\brP_{AB}=\brV_{A}{}^{\brp}\brV_{B\brp}$.   The R-R potential, $\cC^{\alpha}{}_{\bralpha}$, is   set to be an $\Ott$ singlet and  to  assume  a  bi-fundamental  spinorial   representation of  $\Spint\times\oSpint$.  Only after a diagonal gauge fixing of ${\Spint\times\oSpint}$, the DFT-vielbeins and the bi-fundamental R-R potential may be parametrized by the usual   Riemannian variables, \textit{i.e.~}zehnbein, $B$-field and various R-R $p$-form fields. Furthermore, the R-R sector in the diagonal gauge can  be  mapped to an $\Ott$ spinor, as a consequence of     compensating local Lorentz rotation whose job is to  preserve   the diagonal gauge~\cite{Jeon:2012kd}.  It  is an open question  how to couple  the  ${\Spint\times\oSpint}$ bi-fundamental  spinorial  R-R potential to D-branes  as well as  to fundamental strings  in a covariant manner.  \\
\indent \textit{A priori}, $\ODD$ T-duality is not a Noether symmetry of the DFT action, since  the $\ODD$ transformation rotates the section  on which the  theory lives. However, when a dimensional reduction is performed on  flat $n$-dimensional tori, the obvious subgroup, $\mathbf{O}(n,n)$,   becomes  a Noether symmetry (or  the ``enhanced" symmetry) of the reduced action.  DFT   guides how to perform   the $\mathbf{O}(n,n)$ rotations in a convenient way,  which   can be  used as a solution generating technique.  More generically,  T-duality should be able to   act on  any  isometry direction and this procedure    involves  \textit{finite DFT-coordinate transformations}.  Namely,  if the background admits   a generic isometry, but  somehow the given    coordinate system does not manifest it,  in order to apply the DFT $\ODD$  transformation rule, it is  necessary to take the  following steps.  \textit{First} to change the  coordinate system to a new one where  the isometry direction becomes   manifest, \textit{i.e.} all the fields are explicitly independent of one particular  coordinate, \textit{second} to apply the DFT $\ODD$ T-duality  rotation rule, and \textit{third} to come back to the original coordinate system.  \\ 

While the infinitesimal DFT-diffeomorphism generated by the generalized Lie derivative is by now well understood,  \textit{e.g.}  in terms of the semi-covariant,             T-geometry~\cite{Jeon:2010rw,Jeon:2011cn,Jeon:2011vx,Jeon:2012kd,Jeon:2011kp}, or others~\cite{Hohm:2011si,Hohm:2012mf},~\cite{Berman:2013uda},   the finite case has been much less studied,  except \cite{Hohm:2012gk}.   Besides the  practical benefit of obtaining   a solution generating technique,    clear geometric understanding of the finite DFT-diffeomorphism should shed light on  ``non-geometric" (or non-Riemannian)   aspects of supergravity  and eventually string theory itself~\cite{Hull:2004in}.  A first step in this direction was taken in \cite{Hohm:2012gk}. \\
~\\
%%%%%%%%%%%%%%%%%%%%%%%%%%%%%%%%%%%%%%%%%%%%%%%%%%%%%%%%%%%%%%%%%%%%%%%%%%%%%%%%%%%%%%%%%%%%%%%%%%%%%%%%%%%%%%%%%%%%%%%%%%%%%%%%%%%%%%%%%%%%%%%%%%%%%%
\noindent\textbf{{Organization.}}    In this paper, being  motivated by the above necessities,   keeping \cite{Hohm:2012gk} as our key reference, we  conduct  research on  the finite  DFT-diffeomorphism.     The organization of the work   is as follows.
\begin{itemize}
\item  In section~\ref{secCGS}, we argue that   in double field theory  subject to a section condition,  the coordinates  do not    represent the   physical points in an   injective manner. A physical point  should      be rather    one-to-one   identified    with   a   `gauge orbit' in the coordinate space.   We  view then   the DFT-diffeomorphism symmetry as an invariance under arbitrary reparametrizations of the gauge orbits. We name the associated  gauge symmetry,   `\textit{coordinate gauge symmetry}'. 
\item  In section~\ref{SECexp},   taking the \textit{coordinate gauge symmetry} into account,  we   study   the exponential maps,   both for the DFT-coordinates and for the DFT-tensors. In particular,  by formally considering   an exponentiation of the generalized Lie derivative,    we derive  a  tensorial  finite transformation rule,  \textit{c.f.} (\ref{TsR}) and (\ref{Rsolution}). We then show  that,  \textit{up to the coordinate gauge symmetry},  an earlier   tensorial  transformation rule  proposed in \cite{Hohm:2012gk}  is consistent with our  formal solution  and hence with  the  exponential map.  
 \item  In section~\ref{SECcanonical}, we choose a specific section, called \textit{canonical section}, to solve the section condition and to reduce the DFT-geometry to Riemannian geometry.  Upon the canonical section, we identify the coordinate gauge symmetry with the $B$-field gauge symmetry. 
\item  In section~\ref{SECsemi}, we  review, from \cite{Jeon:2010rw,Jeon:2011cn},   the semi-covariant derivatives and  the semi-covariant curvatures, along with 
their full covariantization  with  the help of the projectors for  the  `infinitesimal' DFT-diffeomorphism.   We then discuss an  extension  to the  `finite' case. We show, both for our formal tensorial  transformation rule and for the one proposed in \cite{Hohm:2012gk},  the full covariance of the semi-covariant derivatives and curvatures    persists   for  the   `finite' DFT-diffeomorphism too.   
\item In section~\ref{secCON}, we conclude with \textit{summary} and \textit{comments}. 
  \end{itemize}
\newpage
\noindent\textbf{{Convention.}} Unless mentioned explicitly,  our analyses are   fully $\ODD$ covariant and parametrization-independent, without choosing     any specific   section.  When it is done like section~\ref{SECcanonical}, we shall refer to  the    \textit{canonical choice of the section}. This  means,  for the full DFT-coordinates, $x^{A}=(\ty_{\mu},y^{\nu})$,  and for  the DFT-vielbein,     we set  at least locally~\cite{Jeon:2011cn,Jeon:2011vx},  
\be
\ba{ll}
\dis{\frac{\partial\,~}{\partial{\ty}_{\mu}}\equiv 0\,,}\quad~&~\quad
V_{Ap}=\textstyle{\frac{1}{\sqrt{2}}}{{\left(\ba{c} (e^{-1})_{p}{}^{\mu}\\(B+e)_{\nu p}\ea\right)}}\,.
%%%
%%\quad&\quad\brV_{A{\brp}}=\textstyle{\frac{1}{\sqrt{2}}}\left(\ba{c} (\bre^{-1})_{\brp}{}^{\mu}\\(B+\bre)_{\nu{\brp}}\ea\right)\,,
%%%
\ea
 \label{canonisection}
\ee
For example, with the  \textit{canonical choice of the section}~(\ref{canonisection}),  the infinitesimal DFT-diffeomorphism  parameter, $V^{A}=(\lambda_{\mu},\xi^{\nu})$~(\ref{tcL}), is  clearly divided  into  two parts: the  first half, $\lambda_{\mu}$,  for the $B$-field  gauge symmetry and the second half, $\xi^{\nu}$,    for  the Riemannian   diffeomorphism (or the  ordinary Lie derivative), such that  the generalized Lie derivative acting on the DFT-vielbein, $\delta  V_{Ap}=\hcL_{V}V_{Ap}$, gives rise to
\be
\ba{l}
\delta e_{\mu}{}^{p}=\xi^{\sigma}\partial_{\sigma}e_{\mu}{}^{p}+\partial_{\mu}\xi^{\sigma}e_{\sigma}{}^{p}=\cL_{\xi}e_{\mu}{}^{p}\,,\\
\delta B_{\mu\nu}=\xi^{\sigma}\partial_{\sigma}B_{\mu\nu}+\partial_{\mu}\xi^{\sigma}B_{\sigma\nu}+\partial_{\nu}\xi^{\sigma}B_{\mu\sigma}
+\partial_{\mu}\lambda_{\nu}-\partial_{\nu}\lambda_{\mu}=\cL_{\xi}B_{\mu\nu}+\partial_{\mu}\lambda_{\nu}-\partial_{\nu}\lambda_{\mu}\,.
\ea
\label{eB}
\ee
~\\
\noindent For the sake of simplicity we also often adopt a matrix notation  to suppress spacetime indices. In this case, with one exception,  it is always assumed that the matrices carry  the row index at the   south-west corner and the column index at the   north-east corner, \textit{e.g.~}$M_{A}{}^{B}$.   The only  exceptional matrix is the $\ODD$ metric, $\cJ_{AB}$,  given  in (\ref{ODDmetric}).  Further we define \textit{the $\ODD$ conjugation matrix,} $\brM_{A}{}^{B}$, by
\be
\brM:=\cJ M^{t}\cJ^{-1}\,,
\label{ODDconjugation}
\ee
such that
\be
M\brM=1\quad\Longleftrightarrow\quad M\in\ODD\,.
\ee
While the  two matrices, $M$ and $M^{t}$, cannot be multiplied by each other  due  to their conflicting  index structures  (note `\,$M^{t}{}^{A}{}_{B}$\,'),  the product of $M$ and $\brM$ does make sense. \\

\noindent Throughout the paper, all the fields, both DFT-tensors and DFT-diffeomorphism parameters,  are always assumed to satisfy the section condition~(\ref{seccon}).

%%%%%%%%%%%%%%%%%%%%%%%%%%%%%%%%%%%%%%%%%%%%%%%%%%%%%%%%%%%%%%%%%%%%%%%%%%%%%%%%%%%%%%%%%%%%%%%%%%%%%%%%%%%%%%%%%%%%%%%%

\section{`Coordinate gauge symmetry'\label{secCGS}}

Let us consider an arbitrary DFT-tensor, $T_{A_{1}A_{2}\cdots A_{n}}(x)$. A generic (local) shift of the coordinates, 
\be
x^{A}\rightarrow x^{A}+\Delta^{A}(x)\,,
\ee 
gives
\be
T_{A_{1}\cdots A_{n}}(x+\Delta)=T_{A_{1}\cdots A_{n}}(x)+{\sum_{n=1}^{\infty}}\,\frac{1}{n!}\Delta^{B_{1}}\Delta^{B_{2}}\cdots\Delta^{B_{n}}\partial_{B_{1}}\partial_{B_{2}}\cdots\partial_{B_{n}}T_{A_{1}\cdots A_{n}}(x)\,.
\label{expDelta}
\ee
Hence, in particular,   if the superscript index of $\Delta^{A}$  comes directly from  a DFT-coordinate  derivative, $\partial^{A}=\cJ^{-1 AB}\partial_{B}$,  the shift is trivial  due to     the section condition~(\ref{seccon}),
\be
\ba{ll}
T_{A_{1}\cdots A_{n}}(x+\Delta)=T_{A_{1}\cdots A_{n}}(x)\quad&\mbox{if}\quad \Delta^{A}=\phi\partial^{A}\varphi\quad\mbox{for~\,some}~~\phi~~\mbox{and}~~\varphi\,.
\ea
\label{TensorCGS}
\ee
This  simple observation leads  us to  propose    an \textit{equivalence relation for the DFT-coordinates},
\be
x^{A}~\sim~ x^{A}+\phi\partial^{A}\varphi\,.
\label{equiv0}
\ee
That is to say,  the coordinates in double field theory  do not    represent the   physical points in an injective manner.  {A physical point should  be  rather    one-to-one identified   with   a `gauge orbit' in the coordinate space.}  Henceforth, we call this gauge symmetry~(\ref{equiv0}), `\textit{coordinate gauge symmetry}'.\\
~\\
\indent The \textit{coordinate gauge symmetry} is additive,  being Abelian in nature, 
\be
\ba{lll}
x^{A}~\sim~ x^{A}+\phi\partial^{A}\varphi\,,~\quad x^{A}~\sim~ x^{A}+\phi^{\prime}\partial^{A}\varphi^{\prime}\quad&\Longrightarrow&\quad x^{A}~\sim~ x^{A}+\phi\partial^{A}\varphi+\phi^{\prime}\partial^{A}\varphi^{\prime}\,,
\ea
\ee
and hence in general,
\be
x^{A}~\sim~x^{A}+\phi^{i}\partial^{A}\varphi_{i}\,.
\label{equivalence}
\ee
\indent With the \textit{canonical choice of the section}~(\ref{canonisection}),  for arbitrary constants, $c_{\mu}$, if we choose $\phi=1$ and $\varphi=c_{\mu}y^{\mu}$, we note  from (\ref{equiv0}),
\be
(\ty_{\mu}\,,\,y^{\nu})~\sim~ (\ty_{\mu}+c_{\mu}\,,\,y^{\nu})\,.
\ee
That is to say, upon the canonical  section,   the dual winding coordinates, $\ty_{\mu}$, are all non-physical, irrelevant.    The physical gauge orbits extend through the dual  winding  directions, and   the `ordinary' coordinates, $y^{\nu}$,   alone  faithfully represent  the physical points. Of course, this is  quite a  natural picture to be expected   restricted on  each  local chart.\\

\noindent In general,   with nontrivial $\phi$,  upon the canonical section the \textit{coordinate gauge symmetry} can be identified with  the  \textit{$B$-field gauge symmetry}: From (\ref{eB}),  with  the infinitesimal DFT-diffeomorphism parameter, 
\be
\delta x^{A}=V^{A}=\phi\partial^{A}\varphi\equiv(\phi\partial_{\mu}\varphi, 0)\,,
\label{infiniCGSpara}
\ee
the   generalized Lie derivative produces only  the $B$-field gauge transformation, 
\be
\ba{ll}
\delta e_{\mu}{}^{p}=0\,,\quad&\quad
\delta B_{\mu\nu}=\partial_{\mu}(\phi\partial_{\nu}\varphi)-\partial_{\nu}(\phi\partial_{\mu}\varphi)\,.
\ea
\label{infiniCGS}
\ee
Through  exponentiation, we shall show below that  this identification is still valid for `finite'  transformations,  \textit{c.f.~}(\ref{finiteCGS}).\\

%%%%%%%%%%%%%%%%%%%%%%%%%%%%%%%%%%%%%%%%%%%%%%%%%%%%%%%%%%%%%%%%%%%%%%%%%%%%%%%%%%%%%%%%%%%%%%%%%%%%%%%%%%%%%%%%%%%%%%%%
\section{Exponential map\label{SECexp}}
In this section,   paying attention to  the \textit{coordinate gauge symmetry},  we  investigate   the exponential maps for  DFT-coordinates and   DFT-tensors.

%%%%%%%%%%%%%%%%%%%%%%%%%%%%%%%%%%%%%%%%
%%%%%%%%%%%%%%%%%%%%%%%%%%%%%%%%%%%%%%%%
\subsection{Exponential map for DFT-coordinates}
For a given infinitesimal   DFT-diffeomorphism parameter, $V^{A}(x)$,  introducing  a real parameter, $s$,   we define an exponential map for the DFT-coordinates,
\be
x^{A}~~~\longrightarrow~~~x_{s}^{A}\,,
\ee
by letting
\be
\ba{ll}
\dis{\frac{\rmd x_{s}^{A}}{\rmd s}=V^{A}(x_{s})\,,}\quad&\quad \left.x_{s}^{A}\right|_{s=0}=x^{A}\,.
\ea
\label{defexp}
\ee
It follows from
\be
\dis{\frac{\rmd^{n} x_{s}^{A}}{\rmd s^{n}}=\left(V^{B}(x_{s})\frac{\partial~}{\partial x_{s}^{B}}\right)^{n}x^{A}_{s}\,,}
\ee
that the power expansion of $x_{s}^{A}$ in $s$ reads
\be
\dis{x_{s}^{A}=e^{sV^{B}(x)\partial_{B}}x^{A}}=x^{A}+\sum_{n=1}^{\infty}\,\frac{s^{n}}{n!}(V^{B}(x)\partial_{B})^{n-1}V^{A}(x)\,.
\label{powerx}
\ee
Hence,  with (\ref{defexp}), we note 
\be
\dis{\frac{\rmd x_{s}^{A}}{\rmd s}=V^{A}(x_{s})=V^{B}(x)\partial_{B}x_{s}^{A}\,,}
\label{xderiv}
\ee
and consequently, 
\be
\dis{V^{A}(x_{s})\frac{\partial~}{\partial x_{s}^{A}}=V^{A}(x)\frac{\partial~}{\partial x^{A}}\,.}
\label{invdffop}
\ee
Namely it    is $s$-independent.  This result  can be used, \textit{e.g.~}to show that,\footnote{Both the left hand side and the right hand side of (\ref{shift})  satisfy the same differential equation in $s$,  thanks  to (\ref{invdffop}), while they  have  the same initial value at ${s=0}$.}  generically  acting on an arbitrary  DFT-tensor,  the differential operator, $e^{sV^{B}(x)\partial_{B}}$,  \textit{shifts} the arguments from  $x^{A}$ to $x^{A}_{s}$,
\be
e^{sV^{B}(x)\partial_{B}}T_{A_{1}\cdots A_{n}}(x)=T_{A_{1}\cdots A_{n}}(x_{s})\,.
\label{shift}
\ee
~\\
The inverse map is,   from (\ref{powerx}),  (\ref{invdffop}) and (\ref{shift}), 
\be
\dis{x^{A}=\exp\left(-sV^{B}(x_{s})\frac{\partial~~}{\partial x^{B}_{s}}\right)x_{s}^{A}=e^{sV^{C}(x)\partial_{C}}\left(e^{-sV^{B}(x)\partial_{B}}x^{A}\right)\,.}
\label{inverse}
\ee
Further,   from (\ref{powerx}) and  due to the section condition,  replacing the parameter,  $V^{A}$  by $V^{A}+\phi^{i}\partial^{A}\varphi_{i}$, gives 
\be
x_{s}^{A}\quad\longrightarrow\quad x_{s}^{A}+s\phi^{i}\partial^{A}\varphi_{i}\,.
\label{xsequiv}
\ee
Hence the replacement generates  the \textit{coordinate gauge symmetry} (\ref{equivalence}), 
\be
x_{s}^{A}+s\phi^{i}\partial^{A}\varphi_{i}~~\sim~~ x_{s}^{A}\,.
\ee
This  suggests  us to impose  an equivalence relation now for  the infinitesimal DFT-diffeomorphism parameters, 
\be
V^{A}~~\sim~~V^{A}+\phi^{i}\partial^{A}\varphi_{i}\,.
\label{cgs2}
\ee
~\\
\noindent It is useful   to introduce, from (\ref{powerx}),
\be
f_{s}^{A}(x):=x_{s}^{A}(x)-x^{A}=\sum_{n=1}^{\infty}\,\frac{s^{n}}{n!}(V^{B}(x)\partial_{B})^{n-1}V^{A}(x)\,,
\ee
which satisfies the section condition and also,  for  the inverse map~(\ref{inverse}),  
\be
x^{A}(x_{s})=x_{s}^{A}+f_{-s}^{A}(x_{s})=e^{sV^{B}(x)\partial_{B}}\left(x^{A}+f_{-s}^{A}(x)\right).
\label{invf}
\ee
We further set a matrix, $L$,  by
\be
\dis{L_{A}{}^{B}:=\partial_{A} x_{s}^{B}(x)=\delta_{A}{}^{B}+\partial_{A}f_{s}^{B}(x)\,,}
\label{defL}
\ee
and obtain, from (\ref{invf}),
\be
\dis{L^{-1}{}_{A}{}^{B}=\frac{\partial x^{B}}{\partial x_{s}^{A}}=\delta_{A}{}^{B}+
e^{sV^{C}(x)\partial_{C}}\left[\partial_{A}f_{-s}^{B}(x)\right]}\,.
\label{Linv}
\ee
Their $\ODD$ conjugation matrices~(\ref{ODDconjugation}) are then\footnote{Taking the $\ODD$ conjugation and taking the inverse commute.}
\be
\ba{ll}
\brL_{A}{}^{B}(x)=\delta_{A}{}^{B}+\partial^{B}f_{sA}(x)\,,\quad&\quad
\dis{\brL^{-1}{}_{A}{}^{B}(x)=\overline{L^{-1}}_{A}{}^{B}(x)=
\delta_{A}{}^{B}+e^{sV^{C}(x)\partial_{C}}\left[\partial^{B}f_{-sA}(x)\right]}\,.
\ea
\label{defbrL}
\ee
It is worth while to note that  under the \textit{coordinate gauge symmetry} transformation of the DFT-diffeomorphism  parameter~(\ref{cgs2}),
\be
\ba{lll}
V^{A}~~&\longrightarrow&~~V^{A}+\phi^{i}\partial^{A}\varphi_{i}\,,
\ea
\ee  
the matrices, $L$ and $\brL^{-1}$, transform as 
\be
\ba{lll}
L~~&\longrightarrow&~~(1+sK)L=L+sK\,,\\
\brL^{-1}~&\longrightarrow&~~(1-s\brK)\brL^{-1}\,,
\ea
\label{cgsLL}
\ee
where we set  a traceless nilpotent matrix, 
\be
\ba{lll}
K_{A}{}^{B}:=\partial_{A}(\phi^{i}\partial^{B}\varphi_{i})\,,\quad&\quad
\tr K=0\,,\quad&\quad K^{2}=0\,.
\ea
\label{defK}
\ee
These properties ensure that the determinants are invariant   under the \textit{coordinate gauge symmetry},
\be
\ba{lll}
\det L=\det\brL~~&\longrightarrow&~~\det L=\det\brL\,.
\ea
\label{detinv}
\ee

~\\
Similarly for the derivative of the  DFT-diffeomorphism parameter, we let
\be
\ba{ll}
W_{A}{}^{B}(x)=\partial_{A} V^{B}(x)\,,\quad&\quad 
\brW_{A}{}^{B}(x)=\partial^{B}V_{A}(x)\,.
\ea
\label{defWbrW}
\ee
The section condition implies then
\be
\ba{lllll}
\brW W=0\,,
\quad&\quad \brW L=\brW\,,
\quad&\quad \brL W=W\,,
\quad&\quad KL=K\,,
\quad&\quad \brL\brK=\brK\,,\\
\multicolumn{5}{c}{\brL^{-1}L=L+\brL^{-1}-1\,,\quad\quad\quad\quad \brL L^{-1}=\brL+L^{-1}-1\,,}
\ea
\label{brWW}
\ee
and,   acting on an arbitrary DFT-tensor,   
\be
\ba{ll}
\brL_{A}{}^{B}\partial_{B}T_{C_{1}C_{2}\cdots C_{n}}=\partial_{A}T_{C_{1}C_{2}\cdots C_{n}}\,,\quad&\quad
\brL^{-1}{}_{A}{}^{B}\partial_{B}T_{C_{1}C_{2}\cdots C_{n}}=\partial_{A}T_{C_{1}C_{2}\cdots C_{n}}\,.
\ea
\label{brLPhi}
\ee
Further, with
\be
M_{A}{}^{B}:=\partial_{A}f_{s}^{B}\,,
\ee
we have
\be
\ba{lllll}
L=1+M\,,\quad&\quad
\brL M=M\,,\quad&\quad\brM L=\brM\,,\quad&\quad \brM M=0\,,\quad&\quad\brW M=0\,,
\ea
\label{brLM}
\ee
and
\be
\ba{ll}
L\brL=\brL L+M\brM=\brL(1+M\brM)L\,,\quad&\quad
\brL^{-1}L^{-1}=L^{-1}(1-M\brM)\brL^{-1}\,.
\ea
\label{LbrL}
\ee
Hence, if we set
\be
\ba{ll}
F:=\half(L\brL^{-1}+\brL^{-1}L)\,,\quad&\quad
\brF=\cJ F^{t}\cJ^{-1}=\half(L^{-1}\brL+\brL L^{-1})\,,
\ea
\label{defF}
\ee
it follows
\be
\ba{ll}
F\brF=1\,,\quad&\quad  F\in\ODD\,.
\ea
\label{FbrF}
\ee 
In fact,  it is this  $\ODD$ element,   $F$, that  was  proposed   in \cite{Hohm:2012gk}  as   the matrix  representation of a finite DFT-diffeomorphism.\footnote{Our `passive' $\brF_{A}{}^{B}$ corresponds to  $\cF_{A}{}^{B}$  in  \cite{Hohm:2012gk}. See also the footnote\,\ref{footnoteap}.}   In the next subsection (section~\ref{SECexpT}), we shall study   in detail   the relation     between the matrix,  $F$,  and   the exponentiation of the generalized Lie derivative subject to the coordinate gauge symmetry.  For this purpose, in the remaining of this subsection,  we shall  prepare  some useful formulae      related  to $F$.\\
~\\
\noindent\textbf{\underline{Properties of $F$}}\\
~\\
\noindent From (\ref{brWW}),  (\ref{brLPhi}) and
\be
\ba{ll}
(1-\brL^{-1})L^{-1}=1-\brL^{-1}\,,\quad&\quad L-F=\half(L+1)(1-\brL^{-1})\,,
\ea
\ee
for  an arbitrary DFT-tensor,  we note 
\be
\ba{ll}
\partial_{A}T_{C_{1}C_{2}\cdots C_{n}}&=(LL^{-1})_{A}{}^{B}\partial_{B}T_{C_{1}C_{2}\cdots C_{n}}\\
{}&=\left[FL^{-1}+\half(L+1)(1-\brL^{-1})L^{-1}\right]_{A}{}^{B}\partial_{B}T_{C_{1}C_{2}\cdots C_{n}}\\
{}&=(FL^{-1})_{A}{}^{B}\partial_{B}T_{C_{1}C_{2}\cdots C_{n}}\,+\,\half\left[(L+1)(1-\brL^{-1})\right]_{A}{}^{B}\partial_{B}T_{C_{1}C_{2}\cdots C_{n}}\\
{}&=(FL^{-1})_{A}{}^{B}\partial_{B}T_{C_{1}C_{2}\cdots C_{n}}\,.
\ea
\ee
Hence, up to the section condition, we have
\be
\dis{\partial_{A}\,=\,L_{A}{}^{B}\frac{\partial~~}{\partial x_{s}^{B}}\,\equiv\, F_{A}{}^{B}\frac{\partial~~}{\partial x_{s}^{B}}\,.}
\label{partialF}
\ee
In fact, this is an `active'  rederivation of a `passive' result in  \cite{Hohm:2012gk},\footnote{The \textit{transformation of a field}  is `active' and the \textit{change of a coordinate system}  is `passive': $\phi(x)\rightarrow\phi_{s}(x)$ \textit{vs.}   $\phi^{\prime}(x^{\prime})=\phi(x)$. While the former is more relevant to Noether symmetry, the latter is more popular in  general relativity literature. In this work, we mainly focus  on the finite DFT-diffeomorphism through the exponentiation of the generalized Lie derivative, and hence  the  `active' point of view.  For a `passive' analysis, from (\ref{Linv}), we note
\[
\dis{\frac{\partial~~}{\partial x_{s}^{A}}=\partial_{sA}=\partial_{A}+
e^{sV^{C}(x)\partial_{C}}\left[\partial_{A}f_{-s}^{B}(x)\partial_{B}\right]}\,.
\]
This expression then shows that the section condition is preserved under the change of  the DFT-coordinate systems, as discussed in  \cite{Hohm:2012gk},
\[
\ba{ll}
\partial_{sA}\partial_{s}^{A}\Phi(x)=0\,,\quad&\quad\partial_{sA}\Phi_{1}(x)\partial_{s}^{A}\Phi_{2}(x)=0\,.
\ea
\]\label{footnoteap}} \textit{i.e.} $ \frac{\partial~~}{\partial x_{s}^{A}}\equiv \brF_{A}{}^{B}\partial_{B}\,$. \\

\noindent From (\ref{cgsLL}), under  the \textit{coordinate gauge symmetry},  the matrix, $F$, transforms as
\be
\ba{lll}
F~~&\longrightarrow&~~
\left[1+\half s(K-\brK+K\brL-L\brK)\right]F\,.
\ea
\label{cgsF}
\ee
Further, from (\ref{defL}) and (\ref{defK}),  we can rearrange the expression inside  the  parentheses as\footnote{It is also worth while to  note   
\[
\ba{lll}
\multicolumn{3}{l}{
(K-\brK+K\brL-L\brK)F=K(1+\brL^{-1})-(1+L)\brK\brL^{-1}=FL^{-1}(K-\brK+K\brL-L\brK)\brL^{-1}\,,}\\
\half\left[L^{-1}(K-\brK+K\brL-L\brK)\brL^{-1}\right]_{A}{}^{B}=
\partial^{\prime}_{A}\Upsilon^{\prime B}-\partial^{\prime B}\Upsilon^{\prime}_{A}\,,~~&~~
\partial^{\prime}_{A}=L^{-1}{}_{A}{}^{B}\partial_{B}=\frac{\partial\,~}{\partial x_{s}^{A}}\,,~~&~~
\Upsilon^{\prime A}=\Upsilon^{B}\brL^{-1}{}_{B}{}^{A}\,.
\ea
\]}
\be
\ba{ll}
\half\left(K-\brK+K\brL-L\brK\right)_{A}{}^{B}=\partial_{A}\Upsilon^{B}-\partial^{B}\Upsilon_{A}\,,\quad&\quad\Upsilon^{A}=\phi^{i}\partial^{A}\varphi_{i}+\half \phi^{i}\partial^{C}\varphi_{i}\partial^{A}f_{sC}\,,
\ea
\ee
such that  the \textit{coordinate gauge symmetry}  of $F$~(\ref{cgsF}) can be organized  as
\be
\ba{lll}
F~~&\longrightarrow&~~F^{\prime}F\,,
\ea
\label{Fp1}
\ee
where $F^{\prime}$ is   `another  $F$  matrix'  corresponding to the genuine \textit{coordinate gauge symmetry},  $x^{A}\rightarrow x^{A}+s\Upsilon^{A}$, 
\be
\ba{lll}
F^{\prime}=\half(L^{\prime}\brL^{\prime-1}+\brL^{\prime-1}L^{\prime})\,,\quad&\quad L^{\prime}_{A}{}^{B}=\partial_{A}x^{\prime B}\,,\quad&\quad x^{\prime A}=x^{A}+s\Upsilon^{A}\,\sim\,x^{A}\,,
\ea
\label{Fp2}
\ee
which has   the components,
\be
{F^{\prime}_{A}{}^{B}=\delta_{A}{}^{B}+s(\partial_{A}\Upsilon^{B}-\partial^{B}\Upsilon_{A})\,.}
\label{Fp3}
\ee
~\\
\noindent Now, from the definition of the exponential map~(\ref{defexp}), we obtain
\be
\ba{ll}
\dis{\frac{\rmd ~}{\rmd s}L(x)=L(x)W(x_{s})\,,}\quad&\quad
\dis{\frac{\rmd ~}{\rmd s}\brL^{-1}(x)=-\brL(x)\brW(x_{s})\,,}
\ea
\label{dsAB}
\ee
and hence
\be
\dis{\frac{\rmd ~}{\rmd s}\det{L(x)}=\det{L(x)}\times\tr{W(x_{s})}\,.}
\label{dsdet}
\ee
In particular, the right hand sides of (\ref{dsAB}) and (\ref{dsdet}) are given by products of  two  quantities whose arguments are positioned  at two different points, $x$ and $x_{s}$.\\
~\\
\noindent We   introduce a differential operator, 
\be
\delta_{s}:=\dis{\frac{\rmd ~}{\rmd s}-V^{A}(x)\partial_{A}}\,,
\label{deltas}
\ee
satisfying, first of all,
\be
\delta_{s}x_{s}^{A}(x)=0\,.
\label{deltasxz}
\ee
It follows from (\ref{xderiv})  that,  (\ref{dsAB}) and (\ref{dsdet}) are equivalent to 
\be
\ba{ll}
\dis{\delta_{s}L(x)=W(x)L(x)\,,}\quad&\quad
\dis{\delta_{s}\brL^{-1}(x)=-\brW(x)\brL^{-1}(x)\,,}
\ea
\label{deltasR}
\ee
and
\be
\dis{\delta_{s}\det{L(x)}=\det{L(x)}\times\tr{W(x)}\,.}
\label{deltasdet}
\ee
In contrast to (\ref{dsAB}) and (\ref{dsdet}), the right hand sides are now all positioned at the same point, $x$. We may simply write then
\be
\ba{lll}
\dis{\delta_{s}L=WL\,,}\quad&\quad
\dis{\delta_{s}\brL^{-1}=-\brW\brL^{-1}\,,}\quad&\quad
\dis{\delta_{s}\det{L}=\det{L}\,\tr{W}\,.}
\ea
\ee
Further, exponentiating these expressions we get 
\be
\ba{lll}
L=\dis{\exp\left[{s\left(V{\cdot\partial}+W\right)}\right]1\,,}\quad&\quad 
\brL^{-1}=\dis{\exp\left[{s\left(V{\cdot\partial}-\brW\right)}\right]1\,,}\quad&\quad
\det\!L=\dis{\exp\left[{s\left(V{\cdot\partial}+\tr W\right)}\right]1\,.}
\ea
\label{expLbrL}
\ee
Finally, from (\ref{brWW}), (\ref{brLM}), (\ref{LbrL}) and (\ref{FbrF}), we obtain
\be
\left(\delta_{s}F\right)\brF=\half\left(W+W\brL-\brW-L\brW\right)\,,
\ee
and  hence
\be
\ba{ll}
\delta_{s}F=(W-\brW+\Delta)F\,,\quad&\quad \Delta=\half(W\brM-M\brW)\,.
\ea
\label{deltasF}
\ee
Writing explicitly, 
\be
\Delta_{A}{}^{B}=\half\left(\partial_{A}V^{C}\partial^{B}f_{sC}-\partial_{A}f_{s}^{C}\partial^{B}V_{C}\right)=\partial_{A}\left(\half V^{C}\partial^{B}f_{sC}\right)-\partial^{B}\left(\half V^{C}\partial_{A}f_{sC}\right)\,.
\label{DeltaAB}
\ee
We shall come back to this expression in the next subsection.\\

%%%%%%%%%%%%%%%%%%%%%%%%%%%%%%%%%%%%%%%%%%%%%%%%%%%%%%%%%%%%%%%%%%%%%%%%%%%%%%%%%%%%%%%%%%%%%%%%%%%%%%%%%%%%%%%%%%%%%%%%%%%%%%%%%%%%%%%%%%%%%%%%%%%%%%%%%%%%%%%%
%%%%%%%%%%%%%%%%%%%%%%%%%%%%%%%%%%%%%%%%%%%%%%%%%%%%%%%%%%%%%%%%%%%%%%%%%%%%%%%%
\subsection{Exponential maps for DFT-tensors\label{SECexpT}}
In this subsection,  we turn to the exponential maps for  DFT-tensors. In particular,  we shall discuss   two different, yet equivalent up to the coordinate gauge symmetry,   tensorial exponential maps, and  hence two different yet equivalent  tensorial diffeomorphic   transformation rules, \textit{c.f.} (\ref{BothRFtr}).  \\
\indent We start with the following exponential map,
\be
T_{A_{1}A_{2}\cdots A_{n}}(x)~~~\longrightarrow~~~T_{sA_{1}A_{2}\cdots A_{n}}(x)\,,
\label{TTs}
\ee
which is defined by  the exponentiation of the generalized Lie derivative~(\ref{tcL}),   
\be
\ba{ll}
\dis{\frac{\rmd ~}{\rmd s}T_{sA_{1}\cdots A_{n}}(x)
=\hcL_{V(x)}T_{sA_{1}\cdots A_{n}}(x)\,,}\quad&\quad
\left.T_{sA_{1}\cdots A_{n}}(x)\right|_{s=0}=
T_{A_{1}\cdots A_{n}}(x)\,.
\ea
\label{defexpT}
\ee
Here,  the generalized Lie derivative  acts  on  the finitely-transformed DFT-tensor, $T_{sA_{1}\cdots A_{n}}(x)$,  at the point, $x^{A}$, with the local parameter  not $V^{A}(x_{s})$ but $V^{A}(x)$.   Since double field theories are  well understood to be invariant under the infinitesimal   transformation given  by 
the generalized Lie derivative,     the above    exponential map   realizes   a finite   `Noether'  symmetry transformation rule for the  `finite'  DFT-diffeomorphism.\\

\noindent Especially,   for a scalar density we have
\be
\dis{\frac{\rmd ~}{\rmd s}\phi_{s}(x)=V^{A}(x)\partial_{A}\phi_{s}(x)+\omega\partial_{A}V^{A}(x)\phi_{s}(x)\,,}
\ee
or equivalently, with (\ref{deltas}), 
\be
\dis{\delta_{s}\phi_{s}(x)=\omega\partial_{A}V^{A}(x)\phi_{s}(x)\,.}
\ee
This differential equation   has  the following unique solution, from (\ref{invdffop}) and (\ref{deltasdet}),
\be
\dis{\phi_{s}(x)=\left(\det L\right)^{\omega}\phi(x_{s})=\left(\det L\right)^{\omega}\phi\!\left(x+f_{s}(x)\right)\,.}
\ee
This result illustrates   that  the  exponential map indeed corresponds to the `active' ---rather than `passive'---  transformation under the   DFT-diffeomorphism,   and further that after the transformation the scalar, $\phi_{s}(x)$, still satisfies the section condition.\footnote{\textit{c.f.} (\ref{expDelta}) with $\Delta^{A}~\rightarrow~f_{s}^{A}$.} \\
~\\ \newpage
\noindent Now for a generic DFT-tensor (density),  we set
\be
T_{sA_{1}A_{2}\cdots A_{n}}(x)=\left(\det L\right)^{\omega}R_{A_{1}}{}^{B_{1}}R_{A_{2}}{}^{B_{2}}\cdots R_{A_{n}}{}^{B_{n}}
T_{B_{1}B_{2}\cdots B_{n}}(x_{s})\,.
\label{TsR}
\ee
The condition for the exponential map~(\ref{defexpT}) then reduces to
\be
\ba{ll}
\dis{\delta_{s}R=(W-\brW)R\,,}\quad&\quad \left.R\right|_{s=0}=1\,.
\ea
\label{deltasR2}
\ee
Hence, the solution is, in a similar formal  fashion to (\ref{expLbrL}), 
\be
R=\dis{\exp\left[{s\left(V\cdot\partial+W-\brW\right)}\right]1\,.}
\label{Rsolution}
\ee
Clearly, the matrix, $R$,  and hence the transformed DFT-tensor,  $T_{sA_{1}A_{2}\cdots A_{n}}(x)$,  still satisfy  the section condition. However,   it appears hard   to re-express   the solution~(\ref{Rsolution})   in terms of $L_{A}{}^{B}=\partial_{A}x_{s}^{B}$ (and its inverse)  in a compact manner.  \\
~\\
\noindent From
\be
\dis{\delta_{s}\brR=\brR(\brW-W)\,,}
\ee
it is easy to show, like (\ref{FbrF}),
\be
\ba{ll}
\brR R=1\,,\quad&\quad  R\in\ODD\,.
\ea
\label{RbrR}
\ee 
Further, like (\ref{partialF}), up to the section condition, we have
\be
\dis{\partial_{A}\,\equiv\,R_{A}{}^{B}L^{-1}{}_{B}{}^{C}\partial_{C}=R_{A}{}^{B}\frac{\partial~~}{\partial x_{s}^{B}}\,.}
\label{partialR}
\ee
This  equivalence can be proved,  order by order in $s$, using the following `recurrence' relation,
\be
{}\left[\delta_{s}\,,\,(RL^{-1}-1)_{A}{}^{B}\partial_{B}\right]=(W-\brW)_{A}{}^{B}(RL^{-1}-1)_{B}{}^{C}\partial_{C}-(RL^{-1}-1)_{A}{}^{B}\partial_{B}V^{C}\partial_{C}\,,
\label{obopr}
\ee
with the initial data, $R=L=1$ at $s=0$.  Acting on an arbitrary DFT-tensor  which is subject to the section condition,  this differential operator  is trivial.\footnote{Eq.(\ref{obopr}) is a rearrangement of the expression, 
\[
\left[\delta_{s}\,,\,(RL^{-1})_{A}{}^{B}\partial_{B}\right]=(W-\brW)_{A}{}^{B}(RL^{-1})_{B}{}^{C}\partial_{C}-(RL^{-1})_{A}{}^{B}\partial_{B}V^{C}\partial_{C}\,.
\] 
See also (\ref{Rfinite})  and its derivation later. }\\
~\\
\noindent Henceforth,  we  compare the two matrices,  $R$ and  $F$~(\ref{defF})~\cite{Hohm:2012gk}. First of all, from (\ref{FbrF}),  (\ref{partialF}), (\ref{RbrR}) and (\ref{partialR}),  both are $\ODD$ elements and  satisfy  the  `chain rule',  
\be
\dis{\partial_{A}\equiv R_{A}{}^{B}\frac{\partial~~}{\partial x_{s}^{B}}\equiv F_{A}{}^{B}\frac{\partial~~}{\partial x_{s}^{B}}\,.}
\ee  
On the other hand,  from (\ref{deltasF}), (\ref{DeltaAB}) and  (\ref{deltasR2}),   we note  generically   $\Delta_{A}{}^{B}\neq 0$, and  hence they are distinct, 
\be
F_{A}{}^{B}\,\neq\,R_{A}{}^{B}\,.
\ee
However, as for an alternative tensorial  exponential map  to (\ref{defexpT}) and (\ref{TsR}),   if we let    in terms of $F$,
\be
T^{s}_{A_{1}A_{2}\cdots A_{n}}(x):=\left(\det L\right)^{\omega}F_{A_{1}}{}^{B_{1}}F_{A_{2}}{}^{B_{2}}\cdots F_{A_{n}}{}^{B_{n}}
T_{B_{1}B_{2}\cdots B_{n}}(x_{s})\,,
\label{cT}
\ee
we notice from (\ref{deltasF}) and (\ref{DeltaAB})  that  
\textit{the $s$-derivative of it coincides with  a generalized Lie derivative, }
\be
\ba{ll}
\dis{\frac{\rmd ~}{\rmd s}T^{s}_{A_{1}\cdots A_{n}}(x)
=\hcL_{\cV(x)}T^{s}_{A_{1}\cdots A_{n}}(x)\,,}\quad&\quad
\left.T^{s}_{A_{1}\cdots A_{n}}(x)\right|_{s=0}=
T_{A_{1}\cdots A_{n}}(x)\,,
\ea
\label{expcT}
\ee
\textit{where $\cV^{A}(x)$ is  a new infinitesimal DFT-diffeomorphism parameter,}
\be
\cV^{A}(x)=V^{A}(x)+\half V_{B}(x)\partial^{A}f_{s}^{B}(x)\,.
\label{defcV}
\ee
This new parameter  has $s$-dependence and  hence generically  differs from $V^{A}(x)$, except at $s=0$, 
\be
\ba{ll}
\left.\cV^{A}(x)\right|_{s=0}=V^{A}(x)\,,\quad&\quad \left. f_{s}^{B}(x)\right|_{s=0}=0\,.
\ea
\ee
~\\
\noindent  Using the new parameter,  by analogy with (\ref{xderiv}), we may  also   define  another  coordinate exponential map, 
\be
\ba{lll}
x^{A}~\rightarrow~\hat{x}_{s}^{A}(x)\,,\quad&\quad\dis{\frac{\rmd \hat{x}_{s}^{A}}{\rmd s}=\cV^{B}(x)\partial_{B}\hat{x}^{A}_{s}(x)\,,}\quad&\quad \left.\hat{x}_{s}^{A}\right|_{s=0}=x^{A}\,,
\ea
\label{defexp2}
\ee
which satisfies,  like (\ref{invdffop}),
\be
\dis{\frac{\rmd \hat{x}_{s}^{A}}{\rmd s}\frac{\partial~~}{\partial\hat{x}_{s}^{A}}=\cV^{A}(x)\frac{\partial~~}{\partial x^{A}}\,.}
\ee
The $s$-expansion of $\hat{x}_{s}^{A}$ reads
\be
\hat{x}_{s}^{A}=x^{A}+s V^{A}+\half s^{2}\left(V^{B}\partial_{B}V^{A}+\half V_{C}\partial^{A}V^{C}\right)\,+\,\cdots.
\ee
Nevertheless,     crucially,   $\cV^{A}(x)$  belongs to  the same equivalence class as the original parameter,  according to (\ref{cgs2}),
\be
\cV^{A}(x)~~\sim~~V^{A}(x)\,.
\ee
Hence,   $\hat{x}_{s}^{A}$  and $x^{A}_{s}$ are equivalent. They  represent the same physical point,
\be
\dis{x_{s}^{A}(x)~~\sim~~\hat{x}_{s}^{A}(x)\,,}
\label{xequiv}
\ee
such that, for an arbitrary DFT-tensor,  we have
\be
T_{A_{1}A_{2}\cdots A_{n}}(x_{s})=T_{A_{1}A_{2}\cdots A_{n}}(\hat{x}_{s})\,.
\ee
This coordinate equivalence can be shown more explicitly,  order by order in $s$,   using the formula,
\be
\delta_{s}\left(\hat{x}_{s}^{A}-{x}_{s}^{A}\right)=\half V_{C}\partial^{B}f_{s}^{C}\partial_{B}\hat{x}^{A}_{s}~~\sim~~0\,.
\ee
Therefore,  we conclude that, both  matrices, {$F$ and $R$},  represent the same DFT-coordinate transformation, $x^{A}\,\rightarrow\, x^{A}_{s}$, up to the \textit{coordinate gauge symmetry}. Their  only difference  amounts to the additional  $B$-field gauge symmetry, \textit{c.f.~}(\ref{infiniCGS}) and also  (\ref{finiteCGS}) later. That is to say, both the  tensorial   finite transformation rules, or the tensorial exponential maps, (\ref{TsR}) and (\ref{cT}) which we recall:
\be
\ba{llll}
T_{A_{1}A_{2}\cdots A_{n}}(x)\quad&\longrightarrow&\quad T_{sA_{1}A_{2}\cdots A_{n}}(x)=\left(\det L\right)^{\omega}R_{A_{1}}{}^{B_{1}}R_{A_{2}}{}^{B_{2}}\cdots R_{A_{n}}{}^{B_{n}}
T_{B_{1}B_{2}\cdots B_{n}}(x_{s})\,,\\
T_{A_{1}A_{2}\cdots A_{n}}(x)\quad&\longrightarrow&\quad T^{s}_{A_{1}A_{2}\cdots A_{n}}(x)=\left(\det L\right)^{\omega}F_{A_{1}}{}^{B_{1}}F_{A_{2}}{}^{B_{2}}\cdots F_{A_{n}}{}^{B_{n}}
T_{B_{1}B_{2}\cdots B_{n}}(x_{s})\,,
\ea
\label{BothRFtr}
\ee
equally  well realize the  DFT-diffeomorphism symmetry given by   the  exponential maps,
\be
x^{A}~~~\longrightarrow~~~\dis{x_{s}^{A}(x)~~\sim~~\hat{x}_{s}^{A}(x)\,.}
\ee
After all,    from a practical view, what matters is  whether a given transformation rule of fields  gives rise to  a `Noether symmetry' of the action or not.  DFT enjoys the `infinitesimal' Noether symmetry set by the generalized Lie derivative. The  differential formulae, (\ref{defexpT}) and (\ref{expcT}), then ensure that both transformation rules in (\ref{BothRFtr}) separately realize a `finite' Noether symmetry transformation for   DFT-Lagrangian, 
\be
\ba{lll}
\cL_{\DFT}(x)\quad&\longrightarrow&\quad{\det\!\left(\frac{\partial x_{s}}{\partial x}\right)\times\cL_{\DFT}(x_{s})\,.}
\ea
\ee
~\\
\noindent In particular, we have  for the DFT-vielbein,
\be
\ba{lll}
V_{Ap}(x)\quad&\longrightarrow&\quad V_{sAp}(x)=R_{A}{}^{B}V_{Bp}(x_{s})\,,\\
V_{Ap}(x)\quad&\longrightarrow&\quad V^{s}{}_{Ap}(x)=F_{A}{}^{B}V_{Bp}(x_{s})\,,
\ea
\ee
and for the DFT-dilaton,
\be
\ba{lll}
e^{-2d(x)}\quad&\longrightarrow&\quad e^{-2d_{s}(x)}=e^{-2d^{s}(x)}=\det L \,e^{-2d(x_{s})}\,.
\ea
\ee
Having no $\ODD$ vector index,  there is no distinction between $d_{s}(x)$ and $d^{s}(x)$.  Further, from (\ref{detinv}),   this   DFT-dilaton transformation rule   is unaffected by, or  neutral to,   the \textit{coordinate gauge symmetry}. \\

%%%%%%%%%%%%%%%%%%%%%%%%%%%%%%%%%%%%%%%%%%%%%%%%%%%%%%%%%%%%%%%%%%%%%%%%%%%%%%%%%%%%%%%%%%%%%%%%%%%%
\section{Reduction to Riemannian geometry upon canonical section\label{SECcanonical}}
Upon  the canonical  section~(\ref{canonisection}),  with the $\ty$-independent  DFT-diffeomorphism parameter, 
\be
V^{A}(y)=(\lambda_{\mu}(y),\xi^{\nu}(y))\,,
\ee
the exponential map~(\ref{powerx}), $\,x^{A}\rightarrow x_{s}^{A}=(\ty_{s\mu},y_{s}^{\nu})$, reduces to 
\be
\ba{ll}
\ty_{s\mu}=\ty_{\mu}+\tf_{\mu}(y)\,,\quad&\quad\dis{ y_{s}^{\nu}=e^{s\xi^{\lambda}\partial_{\lambda}}y^{\nu}=y^{\nu}+f^{\nu}(y)\,,}
\ea
\ee
where
\be
\ba{ll}
\tf_{\mu}(y)=\dis{\sum_{n=1}^{\infty}\frac{s^{n}}{n!}\left(\xi^{\lambda}(y)\partial_{\lambda}\right)^{n-1}\lambda_{\mu}(y)\,,}\quad&\quad
f^{\nu}(y)=\dis{\sum_{n=1}^{\infty}\frac{s^{n}}{n!}\left(\xi^{\lambda}(y)\partial_{\lambda}\right)^{n-1}\xi^{\nu}(y)\,.}
\ea
\ee
We have then, for (\ref{defL}),  (\ref{defbrL}) and   (\ref{defF}),
\be
\ba{ll}
L=\left(\ba{cc}1&0\\\partial \tf&l\ea\right)\,,\quad\quad&\quad\quad
\brL^{-1}=\left(\ba{cc}l^{-1t}&0\\-(\partial \tf)^{t}l^{-1t}&1\ea\right)\,,
\ea
\label{canonicalL}
\ee
and
\be
\ba{ll}
F=\left(\ba{cc}l^{-1t}&0\\
\left[\partial\hat{f}-(\partial\hat{f})^{t}\right]l^{-1t}&l
\ea\right)\,,\quad&\quad
\brF=F^{-1}=\left(\ba{cc}l^{t}&0\\
-l^{-1}\left[\partial\hat{f}-(\partial\hat{f})^{t}\right]~&l^{-1}
\ea\right)\,,
\ea
\label{canonicalF}
\ee
where  we put   $D\times D$ matrices, $l$ and $\partial \tf$,   along with  a one-form,  $\hat{f}$,
\be
\ba{lll}
 l_{\mu}{}^{\nu}=\partial_{\mu}y_{s}^{\nu}=\delta_{\mu}{}^{\nu}+\partial_{\mu}f^{\nu}\,,\quad&\quad
 (\partial \tf)_{\mu\nu}=\partial_{\mu}\tf_{\nu}\,,\quad&\quad \hat{f}_{\mu}=\half(\tf_{\mu}+l_{\mu}{}^{\nu}\tf_{\nu})\,.
 \ea
 \label{lpf}
 \ee
It follows,  from (\ref{cT}),  that the `finite'   transformation of  the DFT-vielbein~(\ref{canonisection}),
\be
V_{Ap}(x)~~~\longrightarrow~~~F_{A}{}^{B}V_{Bp}(x_{s})\,,
\label{fVtr}
\ee
is equivalent to the  ordinary  Riemannian diffeomorphism plus the $B$-field gauge symmetry,
\be
\ba{lll}
e_{\mu}{}^{p}(y)\quad&\longrightarrow&\quad \partial_{\mu}y_{s}^{\sigma}e_{\sigma}{}^{p}(y_{s})\,,\\
B_{\mu\nu}(y)\quad&\longrightarrow&\quad \partial_{\mu}y_{s}^{\rho}\partial_{\nu}y_{s}^{\sigma}B_{\rho\sigma}(y_{s})+\partial_{\mu}\hat{f}_{\nu}-
\partial_{\nu}\hat{f}_{\mu}\,.
\ea
\label{canonicaleBtr}
\ee
~\\
\noindent Further, the additional  \textit{coordinate gauge symmetry},  $F\,\rightarrow\,F^{\prime}F$ (\ref{Fp1}), (\ref{Fp2}),   modifies    the  $B$-field gauge symmetry   only,  without  changing   the vielbein  transformation,    
\be
\ba{lll}
e_{\mu}{}^{p}(y)\quad&\longrightarrow&\quad \partial_{\mu}y_{s}^{\sigma}e_{\sigma}{}^{p}(y_{s})\,,\\
B_{\mu\nu}(y)\quad&\longrightarrow&\quad \partial_{\mu}y_{s}^{\rho}\partial_{\nu}y_{s}^{\sigma}B_{\rho\sigma}(y_{s})+\partial_{\mu}(\hat{f}_{\nu}+s\tilde{\Upsilon}_{\nu})-
\partial_{\nu}(\hat{f}_{\mu}+s\tilde{\Upsilon}_{\mu})\,.
\ea
\label{finiteCGS}
\ee
This may have  been  anticipated from the covariant expression,  ${F^{\prime}_{A}{}^{B}=\delta_{A}{}^{B}+s(\partial_{A}\Upsilon^{B}-\partial^{B}\Upsilon_{A})}$ in (\ref{Fp3}).  Conclusively,  upon the canonical section, the \textit{coordinate gauge symmetry} is identified with the \textit{$B$-field gauge symmetry}.\\
~~\\
With  (\ref{defWbrW}), explicitly as  
\be
\ba{lll}
W=\left(\ba{cc}0&0\\\partial\lambda&\partial\xi\ea\right)\,,\quad&\quad
\brW=\left(\ba{cc}(\partial\xi)^{t}&0\\(\partial\lambda)^{t}&0\ea\right)\,,\quad&\quad
W-\brW=\left(\ba{cc}-(\partial\xi)^{t}&0\\\partial\lambda-(\partial\lambda)^{t}&\partial\xi\ea\right)\,,
\ea
\ee
and using
\be
\ba{lll}
\delta_{s}\tf_{\mu}(y)=\lambda_{\mu}(y)\,,\quad&\quad\delta_{s}(\partial_{\mu}\tf_{\nu})=\partial_{\mu}\lambda_{\nu}+\partial_{\mu}\xi^{\lambda}\partial_{\lambda}\tf_{\nu}\,,\quad&\quad \delta_{s}l_{\mu}{}^{\nu}=\partial_{\mu}\xi^{\lambda} l_{\lambda}{}^{\nu}\,,
\ea
\ee
it is straightforward to confirm (\ref{deltasR}) and (\ref{deltasF}),
\be
\ba{lll}
\dis{\delta_{s}L=WL\,,}\quad&\quad
\dis{\delta_{s}\brL^{-1}=-\brW\brL^{-1}\,,}\quad&\quad
\delta_{s}F=(W-\brW+\Delta)F\,,
\ea
\ee
where now
\be
\ba{ll}
\Delta=\left(\ba{cc}0&0\\\partial\theta-(\partial\theta)^{t}&0\ea\right)\,,\quad&\quad
\theta_{\mu}=\half(\lambda_{\sigma}\partial_{\mu}f^{\sigma}+\xi^{\sigma}\partial_{\mu}\tf_{\sigma})\,.
\ea
\ee
~\\
~\\
\noindent On the other hand,  with a one-form, $\Omega_{\mu}$,  if we let for (\ref{TsR}), 
\be
R=\left(\ba{cc}l^{-1t}&0\\
\left[\partial\Omega-(\partial\Omega)^{t}\right]l^{-1t}&l
\ea
\right)\,,
\ee
the defining  property   of  $R$~(\ref{deltasR2}) reduces to
\be
(\delta_{s} R)R^{-1}=
\left(\ba{cc}-(\partial\xi)^{t}&0\\
\partial(\delta_{s}\Omega-\partial\xi\Omega)-\left[\partial(\delta_{s}\Omega-\partial\xi\Omega)\right]^{t}&\partial\xi
\ea\right)=\left(\ba{cc}-(\partial\xi)^{t}&0\\
\partial\lambda-(\partial\lambda)^{t}&\partial\xi\ea\right)=W-\brW\,,
\ee
and hence, with  some scalar, $\Phi$,  the one-form should satisfy 
\be
\delta_{s}\Omega_{\mu}=\lambda_{\mu}+\partial_{\mu}\xi^{\nu}\Omega_{\nu}+\partial_{\mu}\Phi\,.
\ee
However,  the solution seems  to admit no algebraic expression  in terms of $\tf_{\mu}$ and $l_{\mu}{}^{\nu}$  in (\ref{canonicalL}).  This    is in contrast  to the case with $F$, \textit{c.f.} (\ref{canonicalF}).  From (\ref{canonicalF}) and (\ref{canonicaleBtr}),   $F$ and $R$ indeed differ by the $B$-field gauge symmetry only, \textit{i.e.\,} $2\partial_{[\mu}\hat{f}_{\nu]}\,$ \textit{v.s.} $\,2\partial_{[\mu}\Omega_{\nu]}$. \\

%%%%%%%%%%%%%%%%%%%%%%%%%%%%%%%%%%%%%%%%%%%%%%%%%%%%%%%%%%%%%%%%%%%%%%%%%%%%%%%%%%%%%%%%%%%%%%%%%%%%%%%%%%%%%%%%%%%%%%%%%%%
\section{Covariance of the semi-covariant derivatives and curvatures \label{SECsemi}}   
The \textit{semi-covariant derivative}, $\,\na_{A}=\partial_{A}+\Gamma_{A}\,$,  was introduced in \cite{Jeon:2010rw,Jeon:2011cn},
\be
\na_{C}T_{A_{1}A_{2}\cdots A_{n}}
:=\partial_{C}T_{A_{1}A_{2}\cdots A_{n}}-\omega\Gamma^{B}{}_{BC}T_{A_{1}A_{2}\cdots A_{n}}+
\sum_{i=1}^{n}\,\Gamma_{CA_{i}}{}^{B}T_{A_{1}\cdots A_{i-1}BA_{i+1}\cdots A_{n}}\,.
\label{semi-covD}
\ee
with the connection,
\be
\ba{ll}
\Gamma_{CAB}=&2\left(P\partial_{C}P\brP\right)_{[AB]}
+2\left({\brP_{[A}{}^{D}\brP_{B]}{}^{E}}-{P_{[A}{}^{D}P_{B]}{}^{E}}\right)\partial_{D}P_{EC}\\
{}&~-\textstyle{\frac{4}{D-1}}\left(\brP_{C[A}\brP_{B]}{}^{D}+P_{C[A}P_{B]}{}^{D}\right)\!\left(\partial_{D}d+(P\partial^{E}P\brP)_{[ED]}\right)\,.
\ea
\label{Gamma}
\ee
This connection is the DFT analogy of  the Christoffel connection in  Riemannian geometry, as  it is the unique solution to the following requirements~\cite{Jeon:2011cn}.
\begin{itemize}
\item The semi-covariant derivative is compatible with the $\ODD$ metric,
\be
\ba{lll}
\nabla_{A}\cJ_{BC}=0~&\Longleftrightarrow&~\Gamma_{CAB}+\Gamma_{CBA}=0\,.
\ea
\label{nacJ}
\ee
\item  The semi-covariant derivative annihilates the whole NS-NS sector, \textit{i.e.}  the DFT-dilaton\footnote{Since $e^{-2d}$ is a scalar density with weight one, $\na_{A}d=-\half e^{2d}\nabla_{A}e^{-2d}=\partial_{A}d+\half \Gamma^{B}{}_{BA}$.} and the pair of projectors~(\ref{projection}),
 \be
\ba{lll}
\nabla_{A}d=0\,,\quad&\quad
\nabla_{A}P_{BC}=0\,,\quad&\quad\nabla_{A}\brP_{BC}=0\,.
\ea
\label{naNSNS}
\ee
\item The cyclic sum of the connection vanishes,
\be
\Gamma_{ABC}+\Gamma_{CAB}+\Gamma_{BCA}=0\,.
\label{cyclic}
\ee
\item Lastly, the connection corresponds to  a kernel of   rank-six projectors,
\be
\ba{ll}
\cP_{CAB}{}^{DEF}\Gamma_{DEF}=0\,,\quad&\quad \bcP_{CAB}{}^{DEF}\Gamma_{DEF}=0\,,
\ea
\label{kernel}
\ee
where 
\be
\ba{ll}
\cP_{CAB}{}^{DEF}=P_{C}{}^{D}P_{[A}{}^{[E}P_{B]}{}^{F]}+\textstyle{\frac{2}{D-1}}P_{C[A}P_{B]}{}^{[E}P^{F]D}\,,\\
\bcP_{CAB}{}^{DEF}=\brP_{C}{}^{D}\brP_{[A}{}^{[E}\brP_{B]}{}^{F]}+\textstyle{\frac{2}{D-1}}\brP_{C[A}\brP_{B]}{}^{[E}\brP^{F]D}\,.
\ea
\label{P6}
\ee
\end{itemize}
In particular, the two symmetric properties, (\ref{nacJ}) and (\ref{cyclic}),   enable us to replace the ordinary derivatives in the definition of the generalized Lie derivative (\ref{tcL}) by the semi-covariant derivatives~\cite{Jeon:2010rw},
\be
\hcL_{V}T_{A_{1}\cdots A_{n}}=V^{B}\na_{B}T_{A_{1}\cdots A_{n}}+\omega\na_{B}V^{B}T_{A_{1}\cdots A_{n}} +\sum_{i=1}^{n}(\na_{A_{i}}V_{B}-\na_{B}V_{A_{i}})T_{A_{1}\cdots A_{i-1}}{}^{B}{}_{A_{i+1}\cdots  A_{n}}\,.
\ee
The rank-six projectors satisfy the projection property, 
\be
\ba{ll}
{\cP_{CAB}{}^{DEF}\cP_{DEF}{}^{GHI}=\cP_{CAB}{}^{GHI}\,,}\quad&\quad{\bcP_{CAB}{}^{DEF}\bcP_{DEF}{}^{GHI}=\bcP_{CAB}{}^{GHI}\,.}
\ea
\ee
Besides,  they  are  symmetric and traceless,
\be
\ba{ll}
{\cP_{CABDEF}=\cP_{DEFCAB}=\cP_{C[AB]D[EF]}\,,}~~&~~{\bcP_{CABDEF}=\bcP_{DEFCAB}=\bcP_{C[AB]D[EF]}\,,} \\
{\cP^{A}{}_{ABDEF}=0\,,~~~~\,P^{AB}\cP_{ABCDEF}=0\,,}~~&~~
{\bcP^{A}{}_{ABDEF}=0\,,~~~~\,\brP^{AB}\bcP_{ABCDEF}=0\,.}
\ea
\label{symP6}
\ee
~\\
Now, under the infinitesimal DFT-coordinate   transformation set by the generalized Lie derivative, the semi-covariant derivative transforms as
\be
\dis{\delta(\na_{C}T_{A_{1}\cdots A_{n}})=
\hcL_{V}(\na_{C}T_{A_{1}\cdots A_{n}})+
\sum_{i=1}^{n}\,2(\cP{+\bcP})_{CA_{i}}{}^{BDEF}
\partial_{D}\partial_{[E}V_{F]}T_{A_{1}\cdots A_{i-1} BA_{i+1}\cdots A_{n}}\,.}
\label{anomalous}
\ee
The sum on the right hand side corresponds to a potentially  anomalous part against  the full covariance. Hence, in general, the semi-covariant derivative is  not necessarily  covariant.\footnote{However, (\ref{nacJ}) and (\ref{naNSNS}) are exceptions as the anomalous terms vanish identically,  thanks to (\ref{symP6}). }   However, since  the anomalous terms are  projected by the rank-six projectors which satisfy the properties in (\ref{symP6}),  it is  in fact  possible to eliminate  them.  Combined with the projectors,   the semi-covariant derivative  ---as the name indicates---  can be converted into   various fully covariant derivatives~\cite{Jeon:2011cn}:
\be
\ba{l}
P_{C}{}^{D}\brP_{A_{1}}{}^{B_{1}}\brP_{A_{2}}{}^{B_{2}}\cdots\brP_{A_{n}}{}^{B_{n}}
\DO_{D}T_{B_{1}B_{2}\cdots B_{n}}\,,\\
\brP_{C}{}^{D}P_{A_{1}}{}^{B_{1}}P_{A_{2}}{}^{B_{2}}\cdots P_{A_{n}}{}^{B_{n}}
\DO_{D}T_{B_{1}B_{2}\cdots B_{n}}\,,\\
P^{AB}\brP_{C_{1}}{}^{D_{1}}\brP_{C_{2}}{}^{D_{2}}\cdots\brP_{C_{n}}{}^{D_{n}}\DO_{A}T_{BD_{1}D_{2}\cdots D_{n}}\,,\\
\brP^{AB}{P}_{C_{1}}{}^{D_{1}}{P}_{C_{2}}{}^{D_{2}}\cdots{P}_{C_{n}}{}^{D_{n}}\DO_{A}T_{BD_{1}D_{2}\cdots D_{n}}\,,\\
P^{AB}\brP_{C_{1}}{}^{D_{1}}\brP_{C_{2}}{}^{D_{2}}\cdots\brP_{C_{n}}{}^{D_{n}}
\DO_{A}\DO_{B}T_{D_{1}D_{2}\cdots D_{n}}\,,\\
\brP^{AB}P_{C_{1}}{}^{D_{1}}P_{C_{2}}{}^{D_{2}}\cdots P_{C_{n}}{}^{D_{n}}
\DO_{A}\DO_{B}T_{D_{1}D_{2}\cdots D_{n}}\,.
\ea
\label{covT}
\ee
In the above,  and in the paper~\cite{Jeon:2011cn}, the full covariance implies that, the  `infinitesimal'  transformation   coincides with  the generalized Lie derivative,  for example,
\be
\delta(P_{C}{}^{D}\brP_{A_{1}}{}^{B_{1}}\brP_{A_{2}}{}^{B_{2}}\cdots\brP_{A_{n}}{}^{B_{n}}
\DO_{D}T_{B_{1}B_{2}\cdots B_{n}})=\hcL_{V}(P_{C}{}^{D}\brP_{A_{1}}{}^{B_{1}}\brP_{A_{2}}{}^{B_{2}}\cdots\brP_{A_{n}}{}^{B_{n}}
\DO_{D}T_{B_{1}B_{2}\cdots B_{n}})\,.
\label{example}
\ee
~\\
We now  turn  to the `finite' DFT-diffeomorphism  and generalize the above infinitesimal  covariance to the finite case.  We do this for  both  tensorial  transformation rules, one from (\ref{TsR}),
\be
\ba{lll}
e^{-2d(x)}~&\longrightarrow&~e^{-2d_{s}(x)}=\det{L}\, e^{-2d(x_{s})}\,,\\
P_{AB}(x)~&\longrightarrow&~P_{sAB}(x)= R_{A}{}^{C}R_{B}{}^{D}P_{CD}(x_{s})\,,\\
T_{A_{1}A_{2}\cdots A_{n}}(x)~&\longrightarrow&~T_{sA_{1}A_{2}\cdots A_{n}}(x)=\left(\det L\right)^{\omega}R_{A_{1}}{}^{B_{1}}R_{A_{2}}{}^{B_{2}}\cdots R_{A_{n}}{}^{B_{n}}
T_{B_{1}B_{2}\cdots B_{n}}(x_{s})\,,
\ea
\label{Rtr}
\ee
and the other from (\ref{cT})~\cite{Hohm:2012gk},
\be
\ba{lll}
e^{-2d(x)}~&\longrightarrow&~e^{-2d^{s}(x)}=\det{L}\, e^{-2d(x_{s})}\,,\\
P_{AB}(x)~&\longrightarrow&~P^{s}_{AB}(x)=F_{A}{}^{C}F_{B}{}^{D}P_{CD}(x_{s})\,,\\
T_{A_{1}A_{2}\cdots A_{n}}(x)~&\longrightarrow&~
T^{s}_{A_{1}A_{2}\cdots A_{n}}(x)=\left(\det L\right)^{\omega}F_{A_{1}}{}^{B_{1}}F_{A_{2}}{}^{B_{2}}\cdots F_{A_{n}}{}^{B_{n}}
T_{B_{1}B_{2}\cdots B_{n}}(x_{s})\,.
\ea
\label{Ftr}
\ee
The finite transformations of the DFT-dilaton and the projection operator   in (\ref{Rtr}) and (\ref{Ftr}) further induce the transformation of the connection through (\ref{Gamma}), and hence the transformation  of the semi-covariant derivative,
\be
\ba{lll}
\na_{A}=\partial_{A}+\Gamma_{A}\quad&\longrightarrow&\quad\na_{sA}=\partial_{A}+\Gamma_{sA}\,,\\
\na_{A}=\partial_{A}+\Gamma_{A}\quad&\longrightarrow&\quad\na^{s}_{A}=\partial_{A}+\Gamma^{s}_{A}\,.
\ea
\ee
The full  `finite'  covariance of the covariant derivatives listed in (\ref{covT})   means then   that, under      `finite' DFT-diffeomorphism,  they follow precisely the same tensorial  
 transformation rule, either  (\ref{Rtr}),
\be
\ba{l}
\left[P_{sC}{}^{D}\brP_{sA_{1}}{}^{B_{1}}\cdots\brP_{sA_{n}}{}^{B_{n}}
\DO_{sD}T_{sB_{1}\cdots B_{n}}\right]\!(x)\\
=\left(\det L\right)^{\omega}
R_{C}{}^{D}R_{A_{1}}{}^{B_{1}}\cdots R_{A_{n}}{}^{B_{n}}
\left[P_{D}{}^{G}\brP_{B_{1}}{}^{E_{1}}\cdots\brP_{B_{n}}{}^{E_{n}}
\DO_{G}T_{E_{1}\cdots E_{n}}\right]\!(x_{s})\,,
\ea
\label{Rfinite}
\ee
or (\ref{Ftr}),
\be
\ba{l}
\left[P^{s}_{C}{}^{D}\brP^{s}_{A_{1}}{}^{B_{1}}\cdots\brP^{s}_{A_{n}}{}^{B_{n}}
\DO^{s}_{D}T^{s}_{B_{1}\cdots B_{n}}\right]\!(x)\\
=\left(\det L\right)^{\omega}
F_{C}{}^{D}F_{A_{1}}{}^{B_{1}}\cdots F_{A_{n}}{}^{B_{n}}
\left[P_{D}{}^{G}\brP_{B_{1}}{}^{E_{1}}\cdots\brP_{B_{n}}{}^{E_{n}}
\DO_{G}T_{E_{1}\cdots E_{n}}\right]\!(x_{s})\,.
\ea
\label{Ffinite}
\ee
Rather than going through   brute force computations, we can verify these two  relations  by noticing    that,  for each formula    the left hand side and the right hand side satisfy the same  differential equation with  the common initial value  at ${s=0\,}$:  By construction,  the $s$-derivative of them  matches with the generalized Lie derivative,  with the parameter, $V^{A}$ (\ref{defexpT}) for (\ref{Rfinite}),  or with the other  parameter, $\cV^{A}$   (\ref{expcT}) for (\ref{Ffinite}). This essentially proves the full  `finite'  covariance. \\ 
~\\
\noindent Similarly,  with the standard `field strength' of the connection, 
\be
R_{CDAB}=\partial_{A}\Gamma_{BCD}-\partial_{B}\Gamma_{ACD}+\Gamma_{AC}{}^{E}\Gamma_{BED}-\Gamma_{BC}{}^{E}\Gamma_{AED}\,, 
\ee
the \textit{semi-covariant curvature} is defined~\cite{Jeon:2011cn},
\be
S_{ABCD}:=\half\left(R_{ABCD}+R_{CDAB}-\Gamma^{E}{}_{AB}\Gamma_{ECD}\right)\,.
\label{Sdef}
\ee
Under the infinitesimal DFT-diffeomorphism, we have
\be
\delta S_{ABCD}=\hcL_{V}S_{ABCD}+ 2\na_{[A}\left((\cP{+\bcP})_{B][CD]}{}^{EFG}\partial_{E}\partial_{[F}V_{G]}\right)+2\na_{[C}\left((\cP{+\bcP})_{D][AB]}{}^{EFG}\partial_{E}\partial_{[F}V_{G]}\right)\,.
\ee
Hence, with the help of the projectors,  the fully covariant curvatures are~\cite{Jeon:2011cn,Jeon:2011sq}\footnote{Up to the section condition, the scalar and the  Ricci curvatures in (\ref{covcurvature}) are  the only (known)  covariant curvatures.  Although there are various other ways to rewrite    them  because of  the section condition (\textit{c.f.}~the Appendix of \cite{Jeon:2012kd}),   it is the scalar  expression in   (\ref{covcurvature}) that enables   the $1.5$ formalism to work   in  the full order supersymmetric extensions of  ${D=10}$ double field theory,  $\,{\cN=1}$~\cite{Jeon:2011sq} and    ${\cN=2}$~\cite{Jeon:2012hp}.\label{foot1p5}} 
\be
\ba{ll}
(P^{AB}P^{CD}-\brP^{AB}\brP^{CD})S_{ACBD}\,,~~~~&~~~~
P_{A}{}^{C}\brP_{B}{}^{D}S^{E}{}_{CED}\,.
\ea
\label{covcurvature}
\ee
~\\
\noindent Under the finite DFT-diffeomorphism, we now have the following covariant transformations,  
\be
\ba{c}
\left[(P_{s}^{AB}P_{s}^{CD}-\brP_{s}^{AB}\brP_{s}^{CD})S_{sACBD}\right](x)=\left[(P^{sAB}P^{sCD}-\brP^{sAB}\brP^{sCD})S^{s}_{ACBD}\right](x)\\
=\left[(P^{AB}P^{CD}-\brP^{AB}\brP^{CD})S_{ACBD}\right](x_{s})\,,
\ea
\label{Sfinite}
\ee
and 
\be
\ba{l}
\left[P_{sA}{}^{C}\brP_{sB}{}^{D}S_{s}^{E}{}_{CED}\right](x)=R_{A}{}^{M}R_{B}{}^{N}\left[P_{M}{}^{C}\brP_{N}{}^{D}S^{E}{}_{CED}\right](x_{s})\,,\\
\left[P^{s}_{A}{}^{C}\brP^{s}_{B}{}^{D}S^{sE}{}_{CED}\right](x)=F_{A}{}^{M}F_{B}{}^{N}\left[P_{M}{}^{C}\brP_{N}{}^{D}S^{E}{}_{CED}\right](x_{s})\,.
\ea
\label{finiteFS}
\ee
~\\
%%%%%%%%%%%%%%%%%%%%%%%%%%%%%%%%%%%%%%%%%%%%%%%%%%%%%%%%%%%%%%%%%%%%%%%%%%%%%%%%%%%%%%%%%%%%%%%%%%%%%%%%%%%%%%%%%%%%%%%%%%%

%\newpage

%%%%%%%%%%%%%%%%%%%%%%%%%%%%%%%%%%%%%%%%%%%%%%%%%%%%%%%%%%%%%%%%%%%%%%%%%%%%%%%%%%%%%%%%%%%%%%%%%%%%%%%%%%%%%%%%%%%%%%%%%%%%%%%%%%%%%%%%%%%%%%%%%%%%%%%%%%%%%%%%%%%%%%%%%%%%%%%%%%%%%%%%%%%%%%%%%%%%%%%%%%%%%%%%%%%%%%%%%%%%%%%%%%%%%%%%%%%%%%%%
\section{Conclusion\label{secCON}}
We summarize our main assertions.
\begin{itemize}
\item  A physical point  in DFT is   one-to-one   identified  with    a   `gauge orbit' in the coordinate space~(\ref{equivalence}),
\be
x^{A}~\sim~x^{A}+\phi^{i}\partial^{A}\varphi_{i}\,.
\label{ConCGS}
\ee
\item  With the  \textit{coordinate gauge symmetry}~(\ref{ConCGS}),  DFT-diffeomorphism symmetry means  an invariance under arbitrary reparametrizations of the gauge orbits. 
\item  While the {coordinate gauge symmetry} does not change the physical point~(\ref{TensorCGS}),  
\be
\ba{ll}
T_{A_{1}\cdots A_{n}}(x)=T_{A_{1}\cdots A_{n}}(x+\Delta)\,,\quad&\quad
\Delta^{A}=\phi^{i}\partial^{A}\varphi_{i}\,,
\ea
\ee
the corresponding tensorial diffeomorphic   transformation rule  may rotate  the $\ODD$ vector indices of  a DFT-tensor. Yet, this rotation merely amounts to  the $B$-field gauge symmetry, (\ref{infiniCGS}), (\ref{finiteCGS}),  upon the \textit{canonical section}~(\ref{canonisection}).

\item The  {coordinate gauge symmetry}  allows more than one tensorial  diffeomorphic transformation rules  which differ from each other by the $B$-field gauge symmetry, such as (\ref{TsR}) and  (\ref{cT})~\cite{Hohm:2012gk},
\be
\ba{llll}
T_{A_{1}A_{2}\cdots A_{n}}(x)~~&\longrightarrow&~~T_{sA_{1}A_{2}\cdots A_{n}}(x)=\left(\det L\right)^{\omega}R_{A_{1}}{}^{B_{1}}R_{A_{2}}{}^{B_{2}}\cdots R_{A_{n}}{}^{B_{n}}
T_{B_{1}B_{2}\cdots B_{n}}(x_{s})\,,\\
T_{A_{1}A_{2}\cdots A_{n}}(x)~~&\longrightarrow&~~T^{s}_{A_{1}A_{2}\cdots A_{n}}(x)=\left(\det L\right)^{\omega}F_{A_{1}}{}^{B_{1}}F_{A_{2}}{}^{B_{2}}\cdots F_{A_{n}}{}^{B_{n}}
T_{B_{1}B_{2}\cdots B_{n}}(x_{s})\,.
\ea
\label{TsRcT}
\ee
The $s$-derivative of each  transformed DFT-tensor coincides with the generalized Lie derivative,  (\ref{defexpT}) and (\ref{expcT}) respectively, 
\be
\ba{ll}
\dis{\frac{\rmd ~}{\rmd s}T_{sA_{1}\cdots A_{n}}(x)
=\hcL_{V(x)}T_{sA_{1}\cdots A_{n}}(x)\,,}\quad&\quad
\dis{\frac{\rmd ~}{\rmd s}T^{s}_{A_{1}\cdots A_{n}}(x)
=\hcL_{\cV(x)}T^{s}_{A_{1}\cdots A_{n}}(x)\,.}
\ea
\ee
Hence both  transformation rules realize   a `finite' Noether symmetry in double field theory.

\item All the    covariantized \textit{semi-covariant derivatives}~(\ref{covT}) and  \textit{semi-covariant curvatures}~(\ref{covcurvature}), from \cite{Jeon:2010rw,Jeon:2011cn},   are  fully covariant under not only the `infinitesimal' but also the  `finite' DFT-diffeomorphism. They   follow    the    finite covariant  transformation rules,  (\ref{TsRcT}),  too.   
\end{itemize}

\noindent Having the   DFT-tensor finite transformation rules at hand, we may explicitly combine the $\ODD$ rotations and  the   DFT-coordinate   transformations.  In particular, we may perform an $\ODD$ T-duality rotation along an arbitrarily given  isometry  direction.\footnote{For  practical purpose, the second   DFT-tensor transformation rule in (\ref{TsRcT}) from \cite{Hohm:2012gk}  seems to be  a more  convenient choice,  as $F$ appears easier to compute than $R$.}  We first change to a new coordinate system (and hence passive $\brF_{A}{}^{B}$) where the isometry direction is manifest and  all the fields are explicitly  independent of one particular coordinate. We apply straightforwardly  the  DFT $\ODD$ transformation rule along the direction with an $\ODD$ matrix, $O_{A}{}^{B}$,  and afterward   come back to the original coordinate system. Schematically, we perform 
\be
\ba{ll}
F\circ O\circ\brF\,,\quad&\quad O\in\ODD\,.
\ea
\ee
The resulting  DFT-tensor transformation, or a generalized $\ODD$  rotation,  is then
\be
\ba{lll}
T_{A_{1}A_{2}\cdots A_{n}}(x)~~&\longrightarrow&~~(FO\brF)_{A_{1}}{}^{B_{1}}(FO\brF)_{A_{2}}{}^{B_{2}}\cdots (FO\brF)_{A_{n}}{}^{B_{n}}
T_{B_{1}B_{2}\cdots B_{n}}(x)\,.
\ea
\ee
Since the intermediate  $\ODD$ rotation acts  on the isometry direction, the argument of the DFT-tensor, $x^{A}$, does not need to be  changed  in the above transformation. Only the $\ODD$ vector indices of the DFT-tensor are rotated by the non-constant  matrix,  $(FO\brF)_{A}{}^{B}$.  Surely, this  matrix can be more  explicitly computed upon the canonical section, \textit{c.f.~}(\ref{canonicalF}). Application to Poisson-Lie T-duality  or non-Abelian T-duality is of interest~\cite{delaOssa:1992vc,Klimcik:1995ux,Klimcik:1995jn}. \\

\noindent The identification of the {coordinate gauge symmetry} with the $B$-field gauge symmetry was done, both infinitesimally (\ref{infiniCGS}) and finitely (\ref{finiteCGS}),   referring to the canonical parametrization of the DFT-vielbein~(\ref{canonisection}).  This    can be further  straightforwardly extended to   the other   `twin' orthogonal DFT-vielbein (\ref{defV})~\cite{Jeon:2011cn,Jeon:2011vx} and also  to the DFT Yang-Mills vector potential~\cite{Jeon:2011kp}, 
\be
\ba{ll}
\brV_{A{\brp}}=\textstyle{\frac{1}{\sqrt{2}}}\left(\ba{c} (\bre^{-1})_{\brp}{}^{\mu}\\(B+\bre)_{\nu{\brp}}\ea\right)\,,\quad&\quad
\cA_{A}=\left(\ba{c}\Phi^{\mu}\\A_{\nu}+B_{\nu\sigma}\Phi^{\sigma}\ea\right)\,.
\ea
\ee
The generalized Lie derivatives of them  lead to  precisely the   same $B$-field transformation rule as (\ref{infiniCGS}), while  the transformations of other component fields, 
$\bre_{\lambda}{}^{\brp}$,  $ A_{\mu}$,  $\Phi^{\nu}$,   are trivial.\\

\noindent Throughout  the analyses,  we have  strictly imposed the section condition or the strong constraint.  The  relaxation of it (\textit{c.f.}~\cite{Geissbuhler:2011mx,Aldazabal:2011nj,Grana:2012rr,Dibitetto:2012rk,Dibitetto:2012xd,Aldazabal:2013mya,Geissbuhler:2013uka,Musaev:2013rq,Berman:2013uda})  is  beyond the scope of the present paper and remains for  future work.\\

\noindent Generalization to the U-duality in $\cM$-theory~\cite{Berman:2011pe,Berman:2011kg,Berman:2011cg,Malek:2012pw,Berman:2012uy,Berman:2012vc,Musaev:2013rq,Park:2013gaj,Cederwall:2013naa,Cederwall:2013oaa,Godazgar:2013rja}  (or the ``exceptional field theory"~\cite{Cederwall:2013naa,Cederwall:2013oaa})  is also of interest.  In terms of a U-duality invariant tensor, the \textit{coordinate gauge symmetry} will assume the form,
\be
\ba{lll}
x^{A}~&\sim&~x^{A}\,+\,Y^{AB}{}_{CD}\phi^{CDi}\partial_{B}\varphi_{i}\,.
\ea
\ee
Especially, for the $\SLf$ U-geometry~\cite{Park:2013gaj},
\be
\ba{lll}
x^{ab}~&\sim&~x^{ab}\,+\,\epsilon^{abcde}\phi^{i}_{c}{}\partial_{de}\varphi_{i}\,.
\ea
\ee 
~\\
~\\
%%%%%%%%%%%%%%%%%%%%%%%%%%%%%%%%%%%%%%%%%%%%%%%%%%%%%%%%%%%%%%%%%%%%%%%%%%%%%%%%%%%%%%%%%%%%%%%%%%%%%%%%%%%%%%%%%%%%%%%%%%%%%%%%%%%%%%%%%%%%%%%%%%%%%%%%%%%%%%%%%%%%%%%%%%%%%%%%%%%%%%%%%%%%%%%%%%%%%%%%%%%%%%%%%%%%%%%%%%%%%%%%%%%%%%%%%%%%%%%%
\section*{Acknowledgements} 
The author wishes to thank Imtak Jeon for helpful  discussions at various stages of writing up this manuscript, and Clare Hall/DAMTP  for  kind  hospitality  during his  sabbatical visit.  The work was supported by the National Research Foundation of Korea and  the Ministry of Education, Science and Technology with the Grant   No. 2012R1A2A2A02046739,    No.  2010-0002980 and  No.  2005-0049409 (CQUeST).  \\~\\

%%%%%%%
%%%No. 2012R1A6A3A03040350,
%%%%
%\newpage
%%%

%%%%%%%%%%%%%%%%%%%%%%%%%%%%%%%%%%%%%%%%%%%%%%%%%%%%%%%%%%%
%%%%%%%%%%%%%%%%%%%%%%%%%%%%%%%%%%%%%%%%%%%%%%%%%%%%%%%%%%%%%


\newpage






  
\begin{thebibliography}{99}
%\cite{Duff:1989tf}
\bibitem{Duff:1989tf}
  M.~J.~Duff,
  %``Duality Rotations In String Theory,''
  Nucl.\ Phys.\ B {\bf 335} (1990) 610.
  %%CITATION = NUPHA,B335,610;%%
  %125 citations counted in INSPIRE as of 08 Apr 2013


%\cite{Tseytlin:1990nb}
\bibitem{Tseytlin:1990nb}
  A.~A.~Tseytlin,
  %``Duality Symmetric Formulation Of String World Sheet Dynamics,''
  Phys.\ Lett.\  B {\bf 242}, 163 (1990).
  %%CITATION = PHLTA,B242,163;%%

%\cite{Tseytlin:1990va}
\bibitem{Tseytlin:1990va}
  A.~A.~Tseytlin,
  %``Duality Symmetric Closed String Theory And Interacting Chiral Scalars,''
  Nucl.\ Phys.\  B {\bf 350}, 395 (1991).
  %%CITATION = NUPHA,B350,395;%%



%\cite{Siegel:1993xq}
\bibitem{Siegel:1993xq}
  W.~Siegel,
  %``Two vierbein formalism for string inspired axionic gravity,''
  Phys.\ Rev.\  D {\bf 47}, 5453 (1993).
%  [arXiv:hep-th/9302036].
  %%CITATION = PHRVA,D47,5453;%%

%\cite{Siegel:1993th}
\bibitem{Siegel:1993th}
  W.~Siegel,
  %``Superspace duality in low-energy superstrings,''
  Phys.\ Rev.\  D {\bf 48}, 2826 (1993).  
 %[arXiv:hep-th/9305073].
  %%CITATION = PHRVA,D48,2826;%%
  


%\cite{Hull:2009mi}
\bibitem{Hull:2009mi}
  C.~Hull and B.~Zwiebach,
  %``Double Field Theory,''
  JHEP {\bf 0909}, 099 (2009).
%  [arXiv:0904.4664 [hep-th]].
  %%CITATION = JHEPA,0909,099;%%

%\cite{Hull:2009zb}
\bibitem{Hull:2009zb}
  C.~Hull and B.~Zwiebach,
  %``The gauge algebra of double field theory and Courant brackets,''
  JHEP {\bf 0909}, 090 (2009).
%  [arXiv:0908.1792 [hep-th]].
  %%CITATION = JHEPA,0909,090;%%
  
  %\cite{Hohm:2010jy}
\bibitem{Hohm:2010jy}
  O.~Hohm, C.~Hull and B.~Zwiebach,
  %``Background independent action for double field theory,''
  JHEP {\bf 1007}, 016 (2010)  [arXiv:1003.5027 [hep-th]].
  %%CITATION = JHEPA,1007,016;%%



%\cite{Hohm:2010pp}
\bibitem{Hohm:2010pp}
  O.~Hohm, C.~Hull and B.~Zwiebach,
  %``Generalized metric formulation of double field theory,''
  JHEP {\bf 1008}, 008 (2010) [arXiv:1006.4823 [hep-th]].
  %%CITATION = JHEPA,1008,008;%%
  
  
  

  
         
  
%\cite{Jeon:2010rw}
\bibitem{Jeon:2010rw}
  I.~Jeon, K.~Lee and J.-H.~Park,
  %``Differential geometry with a projection: Application to double field
  %theory,''
  JHEP {\bf 1104} (2011) 014.
  [arXiv:1011.1324 [hep-th]].
  %%CITATION = JHEPA,1104,014;%%
  



%\cite{Jeon:2011cn}
\bibitem{Jeon:2011cn}
  I.~Jeon, K.~Lee and  J.-H.~Park,
  %``Stringy differential geometry, beyond Riemann,''
  Phys.\ Rev.\  D {\bf 84 } (2011)  044022  [arXiv:1105.6294 [hep-th]].



  %\cite{Jeon:2011vx}
\bibitem{Jeon:2011vx}
  I.~Jeon, K.~Lee and J.-H.~Park,
  %``Incorporation of fermions into double field theory,''
 JHEP {\bf 11}   (2011)   025 [arXiv:1109.2035 [hep-th]].
  %%CITATION = ARXIV:1109.2035;%%


%\cite{Jeon:2012kd}
\bibitem{Jeon:2012kd}
  I.~Jeon, K.~Lee and J.-H.~Park,
  %``Ramond-Ramond Cohomology and O(D,D) T-duality,''
  JHEP {\bf 1209} (2012) 079
  [arXiv:1206.3478 [hep-th]].
  %%CITATION = ARXIV:1206.3478;%%

%\cite{Jeon:2011kp}
\bibitem{Jeon:2011kp}
  I.~Jeon, K.~Lee and  J.-H.~Park,
  %``Double field formulation of Yang-Mills theory,''
  Phys.\ Lett.\  {\bf B701 } (2011)  260-264 [arXiv:1102.0419 [hep-th]].


  
  

%\cite{Jeon:2011sq}
\bibitem{Jeon:2011sq}
  I.~Jeon, K.~Lee and J.-H.~Park,
  %``Supersymmetric Double Field Theory: Stringy Reformulation of Supergravity,''
  Phys.\ Rev. D  Rapid comm. {\bf 85}  (2012) 081501
  [arXiv:1112.0069 [hep-th]].
  %%CITATION = ARXIV:1112.0069;%%
  
    


   %\cite{Jeon:2012hp}
\bibitem{Jeon:2012hp}
  I.~Jeon, K.~Lee, J.-H.~Park and Y.~Suh,
  %``Stringy Unification of Type IIA and IIB Supergravities under N=2 D=10 Supersymmetric Double Field Theory,''
  arXiv:1210.5078 [hep-th].
  %%CITATION = ARXIV:1210.5078;%%
  
  
  
%\cite{Park:2012xn}
\bibitem{Park:2012xn}
  J.-H.~Park,
  %``Stringy differential geometry for double field theory, beyond Riemann,''
  Phys.\ Part.\ Nucl.\  {\bf 43} (2012) 635.
  %%CITATION = PPNUE,43,635;%%



%\cite{Hohm:2010xe}
\bibitem{Hohm:2010xe}
  O.~Hohm and S.~K.~Kwak,
  %``Frame-like Geometry of Double Field Theory,''
  J.\ Phys.\ A {\bf 44} (2011) 085404
  [arXiv:1011.4101 [hep-th]].
  %%CITATION = ARXIV:1011.4101;%%
  %46 citations counted in INSPIRE as of 08 Apr 2013

  

%\cite{Hohm:2011ex}
\bibitem{Hohm:2011ex}
  O.~Hohm and S.~K.~Kwak,
  %``Double Field Theory Formulation of Heterotic Strings,''
  JHEP {\bf 1106} (2011) 096
  [arXiv:1103.2136 [hep-th]].
  %%CITATION = ARXIV:1103.2136;%%
  %29 citations counted in INSPIRE as of 08 Apr 2013





%\cite{Thompson:2011uw}
\bibitem{Thompson:2011uw}
  D.~C.~Thompson,
  %``Duality Invariance: From M-theory to Double Field Theory,''
  JHEP {\bf 1108} (2011) 125
  [arXiv:1106.4036 [hep-th]].
  %%CITATION = ARXIV:1106.4036;%%
  %29 citations counted in INSPIRE as of 08 Apr 2013

%\cite{Copland:2011yh}
\bibitem{Copland:2011yh}
  N.~B.~Copland,
  %``Connecting T-duality invariant theories,''
  Nucl.\ Phys.\ B {\bf 854} (2012) 575
  [arXiv:1106.1888 [hep-th]].
  %%CITATION = ARXIV:1106.1888;%%
  %21 citations counted in INSPIRE as of 08 Apr 2013

%\cite{Hohm:2011zr}
\bibitem{Hohm:2011zr}
  O.~Hohm, S.~K.~Kwak and B.~Zwiebach,
  %``Unification of Type II Strings and T-duality,''
  Phys.\ Rev.\ Lett.\  {\bf 107} (2011) 171603
  [arXiv:1106.5452 [hep-th]].
  %%CITATION = ARXIV:1106.5452;%%
  %27 citations counted in INSPIRE as of 08 Apr 2013
  
%\cite{Hohm:2011dv}
\bibitem{Hohm:2011dv}
  O.~Hohm, S.~K.~Kwak and B.~Zwiebach,
  %``Double Field Theory of Type II Strings,''
  JHEP {\bf 1109} (2011) 013
  [arXiv:1107.0008 [hep-th]].
  %%CITATION = ARXIV:1107.0008;%%
  %35 citations counted in INSPIRE as of 08 Apr 2013

%\cite{Albertsson:2011ux}
\bibitem{Albertsson:2011ux}
  C.~Albertsson, S.~-H.~Dai, P.~-W.~Kao and F.~-L.~Lin,
  %``Double Field Theory for Double D-branes,''
  JHEP {\bf 1109} (2011) 025
  [arXiv:1107.0876 [hep-th]].
  %%CITATION = ARXIV:1107.0876;%%
  %19 citations counted in INSPIRE as of 08 Apr 2013



%\cite{Hohm:2011cp}
\bibitem{Hohm:2011cp}
  O.~Hohm and S.~K.~Kwak,
  %``Massive Type II in Double Field Theory,''
  JHEP {\bf 1111} (2011) 086
  [arXiv:1108.4937 [hep-th]].
  %%CITATION = ARXIV:1108.4937;%%
  %24 citations counted in INSPIRE as of 08 Apr 2013




%\cite{Kan:2011vg}
\bibitem{Kan:2011vg}
  N.~Kan, K.~Kobayashi and K.~Shiraishi,
  %``Equations of Motion in Double Field Theory: From particles to scale factors,''
  Phys.\ Rev.\ D {\bf 84} (2011) 124049
  [arXiv:1108.5795 [hep-th]].
  %%CITATION = ARXIV:1108.5795;%%
  %9 citations counted in INSPIRE as of 08 Apr 2013


%\cite{Geissbuhler:2011mx}
\bibitem{Geissbuhler:2011mx}
  D.~Geissbuhler,
  %``Double Field Theory and N=4 Gauged Supergravity,''
  JHEP {\bf 1111} (2011) 116
  [arXiv:1109.4280 [hep-th]].
  %%CITATION = ARXIV:1109.4280;%%
  %23 citations counted in INSPIRE as of 08 Apr 2013

%\cite{Aldazabal:2011nj}
\bibitem{Aldazabal:2011nj}
  G.~Aldazabal, W.~Baron, D.~Marques and C.~Nunez,
  %``The effective action of Double Field Theory,''
  JHEP {\bf 1111} (2011) 052
   [Erratum-ibid.\  {\bf 1111} (2011) 109]
  [arXiv:1109.0290 [hep-th]].
  %%CITATION = ARXIV:1109.0290;%%
  %39 citations counted in INSPIRE as of 08 Apr 2013



%\cite{Grana:2012rr}
\bibitem{Grana:2012rr}
  M.~Grana and D.~Marques,
  %``Gauged Double Field Theory,''
  JHEP {\bf 1204} (2012) 020
  [arXiv:1201.2924 [hep-th]].
  %%CITATION = ARXIV:1201.2924;%%
  %21 citations counted in INSPIRE as of 08 Apr 2013


%\cite{Dibitetto:2012rk}
\bibitem{Dibitetto:2012rk}
  G.~Dibitetto, J.~J.~Fernandez-Melgarejo, D.~Marques and D.~Roest,
  %``Duality orbits of non-geometric fluxes,''
  Fortsch.\ Phys.\  {\bf 60} (2012) 1123
  [arXiv:1203.6562 [hep-th]].
  %%CITATION = ARXIV:1203.6562;%%
  %15 citations counted in INSPIRE as of 08 Apr 2013
  
  
%\cite{Dibitetto:2012xd}
\bibitem{Dibitetto:2012xd}
  G.~Dibitetto,
  %``Gauged Supergravities and the Physics of Extra Dimensions,''
  arXiv:1210.2301 [hep-th].
  %%CITATION = ARXIV:1210.2301;%%
  
  
%\cite{Aldazabal:2013mya}
\bibitem{Aldazabal:2013mya}
  G.~Aldazabal, M.~Grana, D.~Marques and J.~A.~Rosabal,
  %``Extended geometry and gauged maximal supergravity,''
  arXiv:1302.5419 [hep-th].
  %%CITATION = ARXIV:1302.5419;%%
  %3 citations counted in INSPIRE as of 08 Apr 2013
  
%\cite{Geissbuhler:2013uka}
\bibitem{Geissbuhler:2013uka}
  D.~Geissbuhler, D.~Marques, C.~Nunez and V.~Penas,
  %``Exploring Double Field Theory,''
  arXiv:1304.1472 [hep-th].
  %%CITATION = ARXIV:1304.1472;%%



%\cite{Copland:2011wx}
\bibitem{Copland:2011wx}
  N.~B.~Copland,
  %``A Double Sigma Model for Double Field Theory,''
  JHEP {\bf 1204} (2012) 044
  [arXiv:1111.1828 [hep-th]].
  %%CITATION = ARXIV:1111.1828;%%
  %12 citations counted in INSPIRE as of 08 Apr 2013

%\cite{Hohm:2011nu}
\bibitem{Hohm:2011nu}
  O.~Hohm and S.~K.~Kwak,
  %``N=1 Supersymmetric Double Field Theory,''
  JHEP {\bf 1203} (2012) 080
  [arXiv:1111.7293 [hep-th]].
  %%CITATION = ARXIV:1111.7293;%%
  %18 citations counted in INSPIRE as of 08 Apr 2013



%\cite{Copland:2012zz}
\bibitem{Copland:2012zz}
  N.~B.~Copland,
  %``Connecting T-duality invariant theories,''
  J.\ Phys.\ Conf.\ Ser.\  {\bf 343} (2012) 012025.
  %%CITATION = 00462,343,012025;%%
  %1 citations counted in INSPIRE as of 08 Apr 2013

%\cite{Kan:2012nf}
\bibitem{Kan:2012nf}
  N.~Kan, K.~Kobayashi and K.~Shiraishi,
  %``Equations of motion in Double Field Theory: from classical particles to quantum cosmology,''
  arXiv:1201.6023 [hep-th].
  %%CITATION = ARXIV:1201.6023;%%
  %1 citations counted in INSPIRE as of 08 Apr 2013

%\cite{Andriot:2012an}
\bibitem{Andriot:2012an}
  D.~Andriot, O.~Hohm, M.~Larfors, D.~Lust and P.~Patalong,
  %``Non-Geometric Fluxes in Supergravity and Double Field Theory,''
  Fortsch.\ Phys.\  {\bf 60} (2012) 1150
  [arXiv:1204.1979 [hep-th]].
  %%CITATION = ARXIV:1204.1979;%%
  %25 citations counted in INSPIRE as of 08 Apr 2013


%\cite{Copland:2012ra}
\bibitem{Copland:2012ra}
  N.~B.~Copland, S.~M.~Ko and J.-H.~Park,
  %``Superconformal Yang-Mills quantum mechanics and Calogero model with OSp(N|2,R) symmetry,''
  JHEP {\bf 1207} (2012) 076
  [arXiv:1205.3869 [hep-th]].
  %%CITATION = ARXIV:1205.3869;%%
  %2 citations counted in INSPIRE as of 08 Apr 2013
  
  

%\cite{Hohm:2011si}
\bibitem{Hohm:2011si}
  O.~Hohm and B.~Zwiebach,
  %``On the Riemann Tensor in Double Field Theory,''
  JHEP {\bf 1205} (2012) 126
  [arXiv:1112.5296 [hep-th]].
  %%CITATION = ARXIV:1112.5296;%%
  %21 citations counted in INSPIRE as of 08 Apr 2013


%\cite{Hohm:2012gk}
\bibitem{Hohm:2012gk}
  O.~Hohm and B.~Zwiebach,
  %``Large Gauge Transformations in Double Field Theory,''
  JHEP {\bf 1302} (2013) 075
  [arXiv:1207.4198 [hep-th]].
  %%CITATION = ARXIV:1207.4198;%%
  %12 citations counted in INSPIRE as of 08 Apr 2013



%\cite{Hohm:2012mf}
\bibitem{Hohm:2012mf}
  O.~Hohm and B.~Zwiebach,
  %``Towards an invariant geometry of double field theory,''
  arXiv:1212.1736 [hep-th].
  %%CITATION = ARXIV:1212.1736;%%
  %7 citations counted in INSPIRE as of 08 Apr 2013



%\cite{Berman:2013uda}
\bibitem{Berman:2013uda}
  D.~S.~Berman, C.~D.~A.~Blair, E.~Malek and M.~J.~Perry,
  %``The O_{D,D} Geometry of String Theory,''
  arXiv:1303.6727 [hep-th].
  %%CITATION = ARXIV:1303.6727;%%
  %1 citations counted in INSPIRE as of 08 Apr 2013

  
            
           
       %%%%%%%%%%% SO FAR DFT %%%%%%%%
       
       
%\cite{Berman:2010is}
\bibitem{Berman:2010is}
  D.~S.~Berman and M.~J.~Perry,
  %``Generalized Geometry and M theory,''
  JHEP {\bf 1106} (2011) 074
  [arXiv:1008.1763 [hep-th]].
  %%CITATION = ARXIV:1008.1763;%%
  %61 citations counted in INSPIRE as of 23 Apr 2013


%\cite{Berman:2011pe}
\bibitem{Berman:2011pe}
  D.~S.~Berman, H.~Godazgar and M.~J.~Perry,
  %``SO(5,5) duality in M-theory and generalized geometry,''
  Phys.\ Lett.\ B {\bf 700} (2011) 65
  [arXiv:1103.5733 [hep-th]].
  %%CITATION = ARXIV:1103.5733;%%
  %42 citations counted in INSPIRE as of 08 Apr 2013


%\cite{Berman:2011kg}
\bibitem{Berman:2011kg}
  D.~S.~Berman, E.~T.~Musaev and M.~J.~Perry,
  %``Boundary Terms in Generalized Geometry and doubled field theory,''
  Phys.\ Lett.\ B {\bf 706} (2011) 228
  [arXiv:1110.3097 [hep-th]].
  %%CITATION = ARXIV:1110.3097;%%
  %21 citations counted in INSPIRE as of 08 Apr 2013
  
  
%\cite{Berman:2011cg}
\bibitem{Berman:2011cg}
  D.~S.~Berman, H.~Godazgar, M.~Godazgar and M.~J.~Perry,
  %``The Local symmetries of M-theory and their formulation in generalised geometry,''
  JHEP {\bf 1201} (2012) 012
  [arXiv:1110.3930 [hep-th]].
  %%CITATION = ARXIV:1110.3930;%%
  %31 citations counted in INSPIRE as of 08 Apr 2013
  
%\cite{Berman:2011jh}
\bibitem{Berman:2011jh}
  D.~S.~Berman, H.~Godazgar, M.~J.~Perry and P.~West,
  %``Duality Invariant Actions and Generalised Geometry,''
  JHEP {\bf 1202} (2012) 108
  [arXiv:1111.0459 [hep-th]].
  %%CITATION = ARXIV:1111.0459;%%
  %30 citations counted in INSPIRE as of 23 Apr 2013



%\cite{Malek:2012pw}
\bibitem{Malek:2012pw}
  E.~Malek,
  %``U-duality in three and four dimensions,''
  arXiv:1205.6403 [hep-th].
  %%CITATION = ARXIV:1205.6403;%%
  %7 citations counted in INSPIRE as of 08 Apr 2013


  
%\cite{Berman:2012uy}
\bibitem{Berman:2012uy}
  D.~S.~Berman, E.~T.~Musaev, D.~C.~Thompson and D.~C.~Thompson,
  %``Duality Invariant M-theory: Gauged supergravities and Scherk-Schwarz reductions,''
  JHEP {\bf 1210} (2012) 174
  [arXiv:1208.0020 [hep-th]].
  %%CITATION = ARXIV:1208.0020;%%
  %15 citations counted in INSPIRE as of 08 Apr 2013

%\cite{Berman:2012vc}
\bibitem{Berman:2012vc}
  D.~S.~Berman, M.~Cederwall, A.~Kleinschmidt and D.~C.~Thompson,
  %``The gauge structure of generalised diffeomorphisms,''
  JHEP {\bf 1301} (2013) 064
  [arXiv:1208.5884 [hep-th]].
  %%CITATION = ARXIV:1208.5884;%%
  %15 citations counted in INSPIRE as of 08 Apr 2013


%\cite{Musaev:2013rq}
\bibitem{Musaev:2013rq}
  E.~T.~Musaev,
  %``Gauged supergravities in 5 and 6 dimensions from generalised Scherk-Schwarz reductions,''
  arXiv:1301.0467 [hep-th].
  %%CITATION = ARXIV:1301.0467;%%
  %3 citations counted in INSPIRE as of 08 Apr 2013


%\cite{Park:2013gaj}
\bibitem{Park:2013gaj}
  J.-H.~Park and Y.~Suh,
  %``U-geometry : SL(5),''
  arXiv:1302.1652 [hep-th].
  %%CITATION = ARXIV:1302.1652;%%
  %6 citations counted in INSPIRE as of 08 Apr 2013



%\cite{Cederwall:2013naa}
\bibitem{Cederwall:2013naa}
  M.~Cederwall, J.~Edlund and A.~Karlsson,
  %``Exceptional geometry and tensor fields,''
  arXiv:1302.6736 [hep-th].
  %%CITATION = ARXIV:1302.6736;%%
  %4 citations counted in INSPIRE as of 08 Apr 2013

%\cite{Cederwall:2013oaa}
\bibitem{Cederwall:2013oaa}
  M.~Cederwall,
  %``Non-gravitational exceptional supermultiplets,''
  arXiv:1302.6737 [hep-th].
  %%CITATION = ARXIV:1302.6737;%%
  %2 citations counted in INSPIRE as of 08 Apr 2013

%\cite{Godazgar:2013rja}
\bibitem{Godazgar:2013rja}
  H.~Godazgar, M.~Godazgar and M.~J.~Perry,
  %``E8 duality and dual gravity,''
  arXiv:1303.2035 [hep-th].
  %%CITATION = ARXIV:1303.2035;%%
  %1 citations counted in INSPIRE as of 08 Apr 2013

%%%%%%%%%%%%%%%%%%%%
%%%%%%%%% SO FAR  U-duality
%%%%%%%%%%%%%%%%%%%%%%%%%%%%%%%%%%%%%%%%


%\cite{Hitchin:2004ut}
\bibitem{Hitchin:2004ut}
  N.~Hitchin,
  %``Generalized Calabi-Yau manifolds,''
  Quart.\ J.\ Math.\ Oxford Ser.\  {\bf 54}, 281 (2003)  [arXiv:math/0209099].
  %%CITATION = QJMAA,54,281;%%

  






%\cite{Hitchin:2010qz}
\bibitem{Hitchin:2010qz}
  N.~Hitchin,
  %``Lectures on generalized geometry,''
  arXiv:1008.0973 [math.DG].
  %%CITATION = ARXIV:1008.0973;%%
  
  
  

%\cite{Gualtieri:2003dx}
\bibitem{Gualtieri:2003dx}
  M.~Gualtieri, Ph.D. Thesis,   Oxford University, 2003.
 % ``Generalized complex geometry,''    
  arXiv:math/0401221.
  %%CITATION = MATH/0401221;%%









  %\cite{Pacheco:2008ps}
\bibitem{Pacheco:2008ps}
  P.~P.~Pacheco and D.~Waldram,
  %``M-theory, exceptional generalised geometry and superpotentials,''
  JHEP {\bf 0809} (2008) 123
  [arXiv:0804.1362 [hep-th]].
  %%CITATION = ARXIV:0804.1362;%%
  
  %\cite{Grana:2008yw}
\bibitem{Grana:2008yw}
M.~Grana, R.~Minasian, M.~Petrini and D.~Waldram,
%``T-duality, Generalized Geometry and Non-Geometric Backgrounds,''
JHEP {\bf 0904} (2009) 075 [arXiv:0807.4527 [hep-th]].
%%CITATION = JHEPA,0904,075;%%  


%\cite{Coimbra:2011nw}
\bibitem{Coimbra:2011nw}
  A.~Coimbra, C.~Strickland-Constable and D.~Waldram,
  %``Supergravity as Generalised Geometry I: Type II Theories,''
  JHEP {\bf 1111} (2011) 091
  [arXiv:1107.1733 [hep-th]].
  %%CITATION = ARXIV:1107.1733;%%
  %40 citations counted in INSPIRE as of 08 Apr 2013
  
  
%\cite{Coimbra:2011ky}
\bibitem{Coimbra:2011ky}
  A.~Coimbra, C.~Strickland-Constable and D.~Waldram,
  %``$E_{d(d)} \Times \Mathbb{R}^+$ Generalised Geometry, Connections and M Theory,''
  arXiv:1112.3989 [hep-th].
  %%CITATION = ARXIV:1112.3989;%%
  %27 citations counted in INSPIRE as of 08 Apr 2013
  
  

  
  %\cite{Coimbra:2012yy}
\bibitem{Coimbra:2012yy}
  A.~Coimbra, C.~Strickland-Constable and D.~Waldram,
  %``Generalised Geometry and type II Supergravity,''
  Fortsch.\ Phys.\  {\bf 60} (2012) 982  [arXiv:1202.3170 [hep-th]].
  %%CITATION = ARXIV:1202.3170;%%        
  
    
   %\cite{Coimbra:2012af}
\bibitem{Coimbra:2012af}
  A.~Coimbra, C.~Strickland-Constable and D.~Waldram,
  %``Supergravity as Generalised Geometry II: $E_{d(d)} \times \mathbb{R}^+$ and M theory,''
  arXiv:1212.1586 [hep-th].
  %%CITATION = ARXIV:1212.1586;%%



      


%\cite{Koerber:2010bx}
\bibitem{Koerber:2010bx}
  P.~Koerber,
  %``Lectures on Generalized Complex Geometry for Physicists,''
  Fortsch.\ Phys.\  {\bf 59} (2011) 169
  [arXiv:1006.1536 [hep-th]].
  %%CITATION = ARXIV:1006.1536;%%
  %29 citations counted in INSPIRE as of 08 Apr 2013

       
       
       %%%%%%%%SO FAR Generalized Geometry%%%%%%%%%%




%\cite{Hull:2004in}
\bibitem{Hull:2004in}
  C.~M.~Hull,
  %``A Geometry for non-geometric string backgrounds,''
  JHEP {\bf 0510} (2005) 065
  [hep-th/0406102].
  %%CITATION = HEP-TH/0406102;%%
  %205 citations counted in INSPIRE as of 12 Apr 2013
  
  
  %\cite{delaOssa:1992vc}
\bibitem{delaOssa:1992vc}
  X.~C.~de la Ossa and F.~Quevedo,
  %``Duality symmetries from nonAbelian isometries in string theory,''
  Nucl.\ Phys.\ B {\bf 403} (1993) 377
  [hep-th/9210021].
  %%CITATION = HEP-TH/9210021;%%
  %151 citations counted in INSPIRE as of 18 Apr 2013

  
  %\cite{Klimcik:1995ux}
\bibitem{Klimcik:1995ux}
  C.~Klimcik and P.~Severa,
  %``Dual nonAbelian duality and the Drinfeld double,''
  Phys.\ Lett.\ B {\bf 351} (1995) 455
  [hep-th/9502122].
  %%CITATION = HEP-TH/9502122;%%
  %116 citations counted in INSPIRE as of 18 Apr 2013


 %\cite{Klimcik:1995jn}
\bibitem{Klimcik:1995jn}
  C.~Klimcik,
  %``Poisson-Lie T duality,''
  Nucl.\ Phys.\ Proc.\ Suppl.\  {\bf 46} (1996) 116
  [hep-th/9509095].
  %%CITATION = HEP-TH/9509095;%%
  %65 citations counted in INSPIRE as of 18 Apr 2013

  \end{thebibliography}
\end{document}